\documentclass[manuscript]{aastex}

\usepackage{amsmath}

\usepackage{xcolor}

\fullcollaborationName{The Friends of AASTeX Collaboration}

\begin{document}

\title{A detailed survey of the parallel mean free path of solar energetic particle protons and electrons}



\author{J.~T. Lang\altaffilmark{1}, R.~D. Strauss\altaffilmark{1,2}, N.~E. Engelbrecht\altaffilmark{1}, J.~P. van den Berg\altaffilmark{1}, N. Dresing\altaffilmark{3}, D. Ruffolo\altaffilmark{4}, and R. Bandyopadhyay\altaffilmark{5}}


\altaffiltext{1}{Center for Space Research, North-West University, Potchefstroom, South Africa} 
\altaffiltext{2}{National Institute for Theoretical and Computational Sciences (NITheCS), South Africa}
\altaffiltext{3}{Department of Physics and Astronomy, University of Turku, Turku, Finland}
\altaffiltext{4}{Department of Physics, Faculty of Science, Mahidol University, Bangkok, Thailand}
\altaffiltext{5}{Department of Astrophysical Sciences, Princeton University, Princeton, USA}

\begin{abstract}

In this work, more than a dozen solar energetic particle (SEP) events are identified where the source region is magnetically well-connected to at least one spacecraft at 1~au. The observed intensity-time profiles, for all available proton and electron energy channels, are compared to results computed using a numerical 1D SEP transport model in order to derive the parallel mean free paths (pMFPs) as a function of energy (or rigidity) at 1~au. These inversion results are then compared to theoretical estimates of the pMFP, using observed turbulence quantities with observationally-motivated variations as input. For protons, a very good comparison between inversion and theoretical results is obtained. It is shown that the observed inter-event variations in the inversion pMFP values can be explained by natural variations in the background turbulence values. For electrons, there is relatively good agreement with pMFPs derived assuming the damping model of dynamical turbulence, although the theoretical values are extremely sensitive to the details of the turbulence dissipation range which themselves display a high level of variation. 

\end{abstract}


\section{Introduction}

The radiation impact of solar energetic particles (SEPs) on space-borne technology and human life is undoubtedly understood to be dangerous. 
SEP transport models are being developed to predict the impact of solar events that inject SEPs into the inner heliosphere, so that warnings can be distributed to anyone \citep{WhitmanEA2023_SEPmodels}. 
SEPs are released into the inner heliosphere after being accelerated during solar eruptions, either from solar flares or coronal mass ejections \citep[CMEs; ][]{Reames1999}. 
The particles are expected to follow the (turbulent) \citet{Parker1958} heliospheric magnetic field (HMF) lines which are carried in Archimedean spirals radially outward by the solar wind (SW). 
However, studies have shown that these particles have been recorded to be longitudinally spread out from the nominal Parker spiral field lines, but with a much lower particle intensity at further angles \citep{PaassiltaEA2018_longitudinalevents}, indicating that some level of perpendicular or cross-field diffusion may also be present \cite[see, e.g.,][]{zhang9,drogeEA14,Straussetal2017a}.

The diffusion of energetic particles can be characterised through the mean free path (MFP) corresponding to a diffusion coefficient associated with diffusion in a particular direction. The perpendicular and parallel MFPs describe the diffusion across the field lines and along the Parker spiral, respectively, with the two combined into a radial MFP describing the diffusion along a radial line from the Sun to the observation point. This study utilises a one-dimensional (1D) transport model to simulate the SEP intensity along the Parker spiral, and as such, the parallel MFP (from now on referred to as the pMFP) is the main parameter that is inferred from the model-observations comparison. By comparing model simulations of SEP time series to appropriate observations, the pMFP can be derived from the {\it in-situ} particle observations \citep[e.g.][]{BieberEA94,Droge00,AguedaEA2014,Ruffoloetal1998}. 
These experimentally-derived results for the pMFP can then be compared to theoretical estimates. This is done here for a large number of SEP events, observed near $1$~au, for both electrons and protons, using all available energy channels for the spacecraft considered. 

This study inverts over a dozen such magnetically well-connected events associated with solar flares from $1998$ to $2022$, using measurements from instruments onboard \textit{in situ} spacecraft. 
The instruments used in this work include the Wind Three-Dimensional Plasma and Energetic Particle Investigation \citep[3DP; ][]{LinEA95_WIND}, the High and Low Energy Detectors (HED and LED) found within the SOHO \citep{domingoEA95_SOHO} Energetic and Relativistic Nuclei and Electron experiment \citep[ERNE;][]{TorstieEA91_erne}, and the Solar Electron Proton Telescope \citep[SEPT;][]{MullerMellinEA08_SEPT_sta} and High Energy Telescope \citep[HET;][]{vonRosenvingeEA08_HET_sta} that are found onboard the STEREO-A spacecraft \citep[STA;][]{kaiser05_STAB}. 

The process for finding each event is described in Section~\ref{sec:findingEvents}.
The SEP transport model used in this study is then briefly discussed in Section~\ref{sec:model}. 
The section ends with a summary of the best-fitting pMFP results as a function of rigidity, which is where results from this study are compared to those from previous studies.

Section~\ref{sec:comparisonWtheory} turns to the analytical estimates for the behaviour of the pMFP in terms of rigidity, originally from \citet{BieberEA94}, further expanded on by \citet{TS02,TS03}, and summarized by \citet{EngelbrechtBurger13}.
Evaluating these theoretical pMFP estimates {requires various turbulence parameter values as inputs}. As these have been observed to vary significantly, including variations as a function of the solar cycle \cite[see, e.g.,][]{IsaacsEA15,OughtonEA15,ZhaoEA18,EW20,CuestaEA22,BurgerEA22,wrenchEA24}, an ensemble of the values of turbulence quantities relevant to the theoretical pMFP expressions considered here is constructed. These values are then used to ascertain the potential variation of pMFPs, and compared with the observed values for this quantity reported on here, including best-fit estimations. The paper concludes with a discussion of the implications of said results.


\section{Event selection} \label{sec:findingEvents} 

\begin{figure}[ht!]
    \centering
    \includegraphics[width=0.49\textwidth]{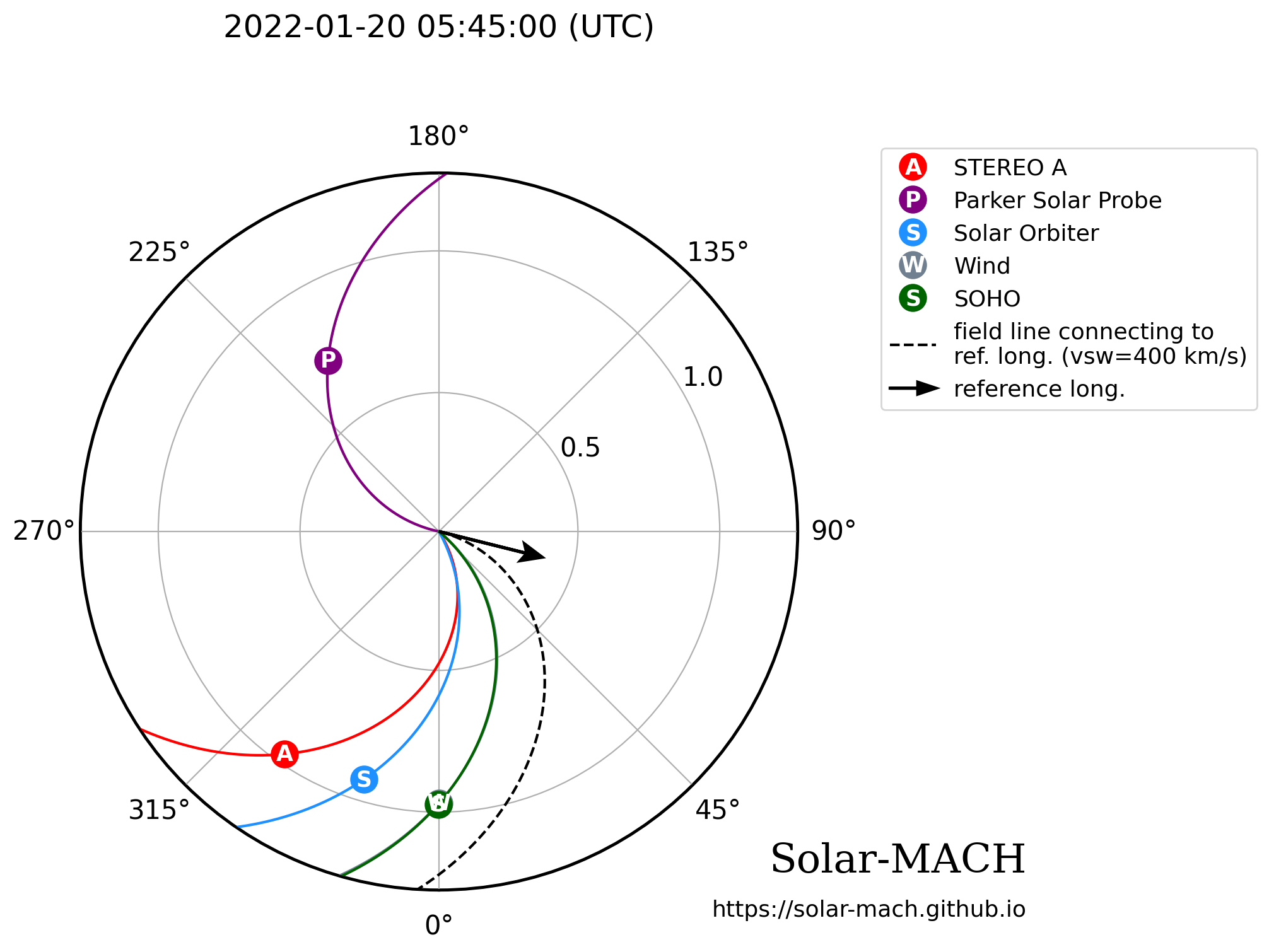}
    \caption{A visual representation of the magnetic connection (using Carrington coordinates) and radial distance of several spacecraft in orbit at 05:45 on 20 January 2022. The flare, occurring at N08W76 in Stonyhurst coordinates, is denoted as the black arrow and the dashed Parker spiral. The reference spirals for each spacecraft are calculated using the SW speed measured onboard, while the spiral of the solar flare is calculated using $V_{sw}=400$~km~s$^{-1}$. This figure was created using the Solar-MACH Python tool \citep{solarMACH}. }
    \label{fig:solarMACH}
\end{figure}

The particle flux-time profiles over various energy ranges from solar event observations over three solar cycles ($1998$~-~$2022$) are used to infer various transport characteristics. 
This is done by fitting the time profile of a predictive SEP transport model to the time profiles of solar events observed by \textit{in situ} instruments onboard spacecraft to infer the pMFP as well as parameters setting the injection profile. 
The model utilized in this work is the SEP propagator model\footnote{Available online at \url{https://github.com/RDStrauss/SEP\_propagator/}} by \citet{vandenBergEA20}. 
This propagation model simulates 1D particle transport from the source surface of the SW and HMF, assumed to be located at a radial distance of $0.05$~au \cite[which is close to the observational range given by][]{Goelzer}, to a given radial distance in the inner heliosphere.
The model restricts the simulations to describe the particle transport along a single Parker HMF line, and as such the model can only be inverted with observations from magnetically well-connected spacecraft. 
This path along the Parker spiral is seen in Figure~\ref{fig:solarMACH} where the colored lines denote the connecting magnetic field lines from the spacecraft to a point on the surface of the Sun. This point is referred to as the magnetic footprint of the spacecraft. The longitudinal distance between the event source and the magnetic footprint is referred to as the longitudinal separation angle $\Delta \Phi$.

For a spacecraft to be considered magnetically well-connected to a solar event, $\Delta \Phi$ must be within range of the source eruption at the time of the event.
In this study, we chose to use a range of $35^{\circ}$ in either direction, {similar to that of \citet{Roelof2015}} who used a maximum separation angle of $30^{\circ}$.
However, \citet{DresingEA2014_stereo_widespread}, which specifically studied longitudinally widespread events, shows that this value can be increased for certain events, as fairly high anisotropies can be seen up to $50^{\circ}$ of longitudinal separation.

\begin{figure}[ht!]
    \centering
    \includegraphics[width=0.49\textwidth]{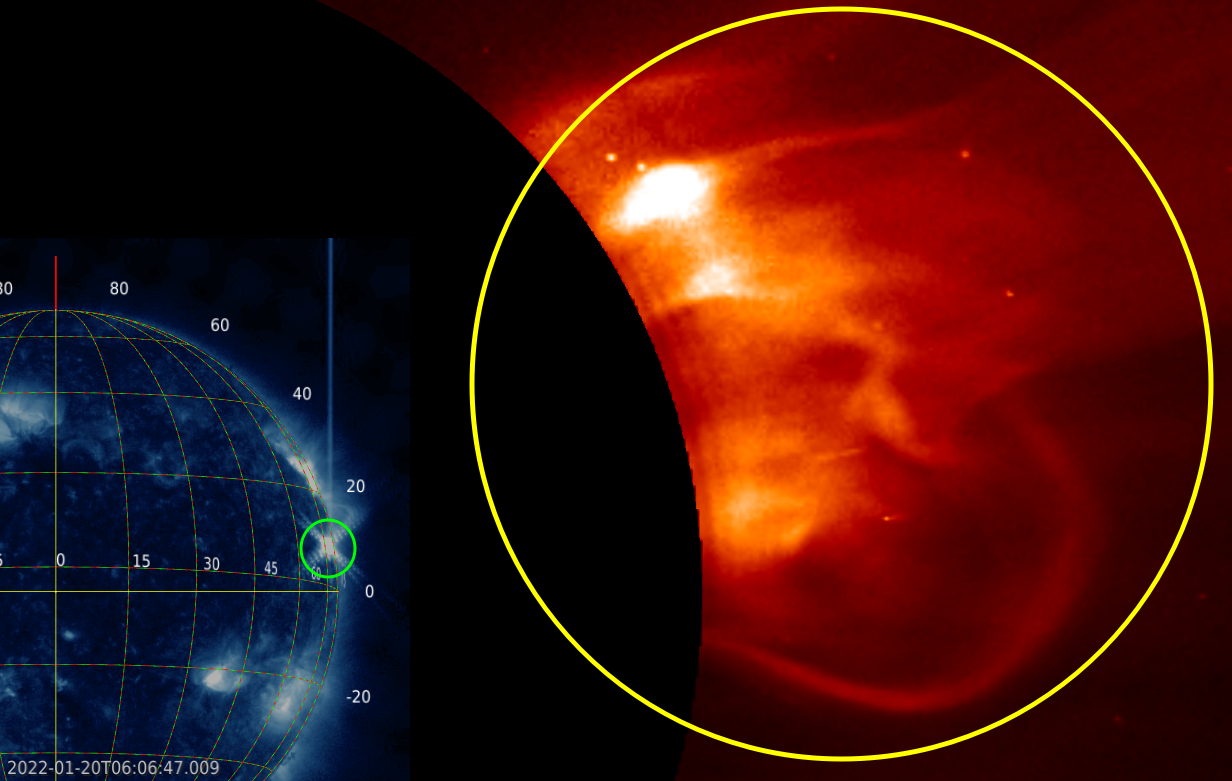}
    \caption{A snapshot taken by SDO/AIA 335 (blue inner image) and SOHO/LASCO C2 (red outer image) of the solar flare (within the green circle) and CME (within the yellow circle) soon after the flare peaked (at 06:06), showing the coordinates at approximately N08W76 in Stonyhurst coordinates. The image was obtained using JHelioviewer \citep{helioviewer}.}
    \label{fig:helioviewer}
\end{figure}

Solar flares are classified based on their soft X-ray peak flux magnitudes from smallest to largest as A, B, C, M, and X \citep{UrsiEA23_AGILEcatalog}, where background solar activity is predominantly found to be within the B-class. 
Combing through multiple online SEP event catalogs to preferably find the larger (M- or X-class) flares, the date, time, and size of the flare (and sometimes the approximate heliographic coordinates too) are recorded. 
Smaller-classed flares can also be used (as seen in Table~\ref{tab:events_fitted} with the B4.2 class flare on 21 March 2022), as long as the intensity displays a clear time profile as a result of the solar flare. 
The JHelioviewer desktop program \citep{helioviewer} is then used to find (or confirm) the heliographic coordinates of the solar event in Stonyhurst or Carrington coordinates. 
JHelioviewer uses a catalog of images taken by {\textit{in situ} {instruments. This includes} the SOHO Large Angle and Spectrometric COronagraph experiments \citep[LASCO C2, and C3;][]{BruecknerEA95_LASCO_soho} and the Solar Dynamics Observatory Atmospheric Imaging Assembly \citep[SDO/AIA;][]{LemenEA12_AIA_sdo},} and overlays them onto a solar surface coordinate system that also corresponds to a solar event catalog called the Heliophysics Event Knowledgebase \citep[HEK;][]{hurlburtEA2012_HEK}, as seen in Figure~\ref{fig:helioviewer}.

The date, time, and approximate heliographic coordinates are then entered into the Solar-MACH Python tool\footnote{Available online at \url{https://solar-mach.github.io/}.} \citep{solarMACH} to obtain the radial distance from the Sun and coordinates (at the radial distance and at the footprint on the Sun) of various helio-spacecraft and planetary bodies relative to the solar event source. 
{We chose to only consider spacecraft near $1$~au with publicly accessible particle intensity data that could be accessed using the \texttt{data\_loader} Python tool designed {by the SERPENTINE Project} \citep[see][ for more details]{palmroosEA2022_SERPENTINEdataloader}, which provided us with Wind, SOHO, and STA data.}
Solar-MACH sources the spacecraft and planetary body data from JPL Horizons\footnote{A solar system database available online at \url{https://ssd.jpl.nasa.gov/horizons/}}, as well as the SW speed $V_{sw}$ used to calculate the Parker spirals seen in Figure~\ref{fig:solarMACH}. 
The tool provides both a visual and numerical output, where Figure~\ref{fig:solarMACH} shows the visual output for a solar event on 20 January 2022 at the start of the eruption. 
This event is well-connected to the Wind and SOHO spacecraft as their footprints are both 26$^{\circ}$ from the source flare, which lies within the 35$^{\circ}$ limit established earlier.

The online solar data collection, Solar Monitor\footnote{\url{https://www.SolarMonitor.org}} \citep{solarmonitor}, utilizing data from the NOAA Space Weather Prediction Center, recorded the M-class flare for this event, peaking at 06:01\footnote{All times are provided in Coordinated Universal Time (UT).}, at approximately N08W76. 
An image of the source region for this event is found within the green circle to the left of Figure~\ref{fig:helioviewer}, alongside a corresponding CME in the yellow circle to the right; thus, confirming the date, time and coordinates provided by Solar Monitor.

{Lastly, the event needed to have produced well-defined intensity-time profiles from at least one magnetically well-connected instrument for it to be considered in this study. This is explained in more detail in Section~\ref{sec:onsetAlignment}.}

\subsection{Catalog of events} \label{sec:catalog}


The catalog of the events measured by instruments roughly $1$~au from the Sun that were studied and fitted in this work is summarised in Table~\ref{tab:events_fitted}, along with the longitudinal separation and radial distance of the observing spacecraft.
The observational data from each mission was accessed using the \texttt{data\_loader} Python tool \citep{palmroosEA2022_SERPENTINEdataloader}.

\begin{deluxetable}{rcCccCC}
\tabletypesize{\scriptsize}
\tablewidth{0pt} 
\tablecaption{Information on each solar event and magnetically connected spacecraft used in the results presented in this study.  \label{tab:events_fitted}}
\tablehead{
\colhead{Date and} & \colhead{GOES} & \colhead{Flare}  & \colhead{} & \colhead{} & \colhead{Radial}\\
\colhead{flare start} & \colhead{flare} & \colhead{location}  & \colhead{Spacecraft} & \colhead{$\Delta \Phi$} & \colhead{distance}\\
\colhead{time} & \colhead{class} & \colhead{(Stonyhurst)}  & \colhead{} & \colhead{[$^{\circ}$]} & \colhead{[au]}
} 
\startdata 
\hline
22-11-1998 & \multirow{2}{*}{X3.7} & \multirow{2}{*}{N27W43} & \multirow{2}{*}{SOHO} & \multirow{2}{*}{21} & \multirow{2}{*}{0.98} \\ 
06:30      &                       &                              &    &    &   \\
\specialrule{.05em}{.05em}{.05em}
11-06-1999 & \multirow{2}{*}{C1.0} & \multirow{2}{*}{S19W44}       &  Wind &  19 & 1.01 \\ 
01:05  &                           &                               &  SOHO &  13 & 1.00 \\
\specialrule{.05em}{.05em}{.05em}
19-10-2001 & \multirow{2}{*}{X1.6} & \multirow{2}{*}{S13W54}   &  Wind &  7 & 0.99 \\ 
00:43      &                       &                           &  SOHO & 23 & 0.98 \\
\specialrule{.05em}{.05em}{.05em}
20-02-2002 & \multirow{2}{*}{M5.1} & \multirow{2}{*}{S11W56}   &  Wind &  5 & 0.98 \\ 
05:42      &                       &                           &  SOHO &  5 & 0.98 \\
\specialrule{.05em}{.05em}{.05em}
01-11-2004 & \multirow{2}{*}{C2.9} & \multirow{2}{*}{N06W48}     &  Wind &  13 & 0.99 \\ 
06:55      &                       &                             &  SOHO &  12 & 0.98 \\
\specialrule{.05em}{.05em}{.05em}
14-08-2010 & \multirow{2}{*}{C4.4} & \multirow{2}{*}{N13W56}    &  Wind &  0 & 1.01 \\ 
09:38      &                       &                            &  SOHO &  1 & 1.00 \\
\specialrule{.05em}{.05em}{.05em}
09-08-2011 & \multirow{2}{*}{X6.9} & \multirow{2}{*}{N16W70}    &  Wind &  -8 & 1.00 \\ 
07:48      &                       &                            &  SOHO & -29 & 1.00 \\
\specialrule{.05em}{.05em}{.05em}
17-05-2012 & \multirow{2}{*}{M5.1} & \multirow{2}{*}{N11W78}    &  Wind &  -16 & 1.00 \\ 
01:25      &                       &                            &  SOHO &   -7 & 1.00 \\
\specialrule{.05em}{.05em}{.05em}
20-09-2015 & \multirow{2}{*}{M2.1} & \multirow{2}{*}{S22W50}    &  Wind &  11 & 0.99 \\ 
17:32      &                       &                            &  SOHO &  -5 & 1.00 \\
\specialrule{.05em}{.05em}{.05em}
09-10-2021 & \multirow{2}{*}{M1.6} & \multirow{2}{*}{N18E08} & \multirow{2}{*}{STA} & \multirow{2}{*}{32} & \multirow{2}{*}{0.96}\\ 
06:19   &                           &       &           &       &       \\
\specialrule{.05em}{.05em}{.05em}
28-10-2021 & \multirow{2}{*}{X1.0} & \multirow{2}{*}{S28W01} & \multirow{2}{*}{STA} & \multirow{2}{*}{21} & \multirow{2}{*}{0.96}\\ 
15:17       &     &       &   &  & \\
\specialrule{.05em}{.05em}{.05em}
\multirow{2}{*}{18-01-2022}& \multirow{3}{*}{M1.5} & \multirow{3}{*}{N08W55} & STA  & -30 & 0.97 \\ 
\multirow{2}{*}{17:01}  &                       &                            & Wind & -14 & 0.97 \\
                        &                       &                            & SOHO & -14 & 0.98 \\
\specialrule{.05em}{.05em}{.05em}
20-01-2022 & \multirow{2}{*}{M5.5} & \multirow{2}{*}{N08W76}  &  Wind &  -26 & 0.97 \\ 
05:41  &                       &                          &  SOHO &  -26 & 0.98 \\
\specialrule{.05em}{.05em}{.05em}
\multirow{2}{*}{21-03-2022} & \multirow{3}{*}{B4.2} & \multirow{3}{*}{{N15W50}} &  STA &  -21 & 0.97 \\ 
\multirow{2}{*}{05:38}  &           &                   &  Wind &  5 & 0.99 \\
                          &           &                   &  SOHO &  5 & 0.99 \\
\specialrule{.05em}{.05em}{.05em}
02-04-2022  & \multirow{2}{*}{M3.9} & \multirow{2}{*}{N12W68} &  Wind &  -16 & 0.99 \\ 
12:56       &                       &                         &  SOHO &  -17 & 0.99 \\
\enddata
\tablecomments{
The flare start time, class, and location were mostly taken from Solar Monitor, with the time and location confirmed using JHelioviewer. 
}
\end{deluxetable}

The longitudinal separation angle $\Delta \Phi$ is a vector measurement of the distance {between the spacecraft's footprint and} the location of the solar event, where the sign is indicative of whether the spacecraft lies to the east or west of the flare.
A positive $\Delta \Phi$ indicates that the spacecraft's footprint is west of the solar event and a negative sign that the footprint is east of the event.

\begin{figure*}[ht!]
    \centering
    \includegraphics[width=0.85\textwidth]{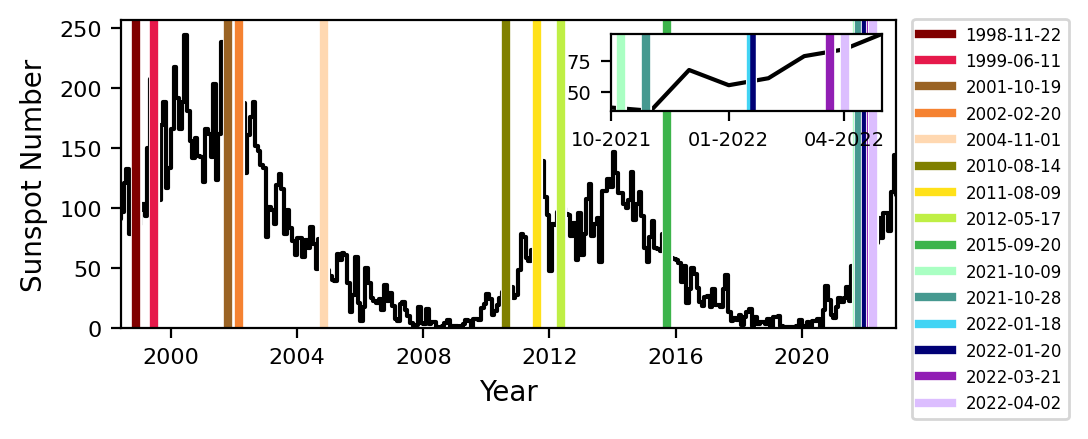}
    \caption{Average yearly sunspot numbers from 1998 to 2022 using data drawn using the SunPy Python module \citep{sunpy}. The colored vertical lines indicate when each of the events that are considered in this paper occurred.}
    \label{fig:solarCycle}
\end{figure*}

Figure~\ref{fig:solarCycle} is provided as a reference for when the events in Table~\ref{tab:events_fitted} occurred in relation to the solar cycle, with observed sunspot number as a proxy for the level of solar activity. 
Each event corresponds to a different colored line, and these same color references are also utilized later in Figure~\ref{fig:mfpVrig}. 
Not unexpectedly, most of the events considered here correspond to periods of intermediate to maximum solar activity.

\subsection{SEP event on 20 January 2022} \label{sec:event}

As a typical example of the SEP fitting procedure used in this work, the SEP event on 20 January 2022 is discussed in detail. 
Just before 06:00 on 20 January 2022, an M5.5-class flare erupted, with an accompanying fast-CME, from near the solar west limb.

\begin{figure*}[ht!]
    \centering
    \includegraphics[width=0.99\textwidth]{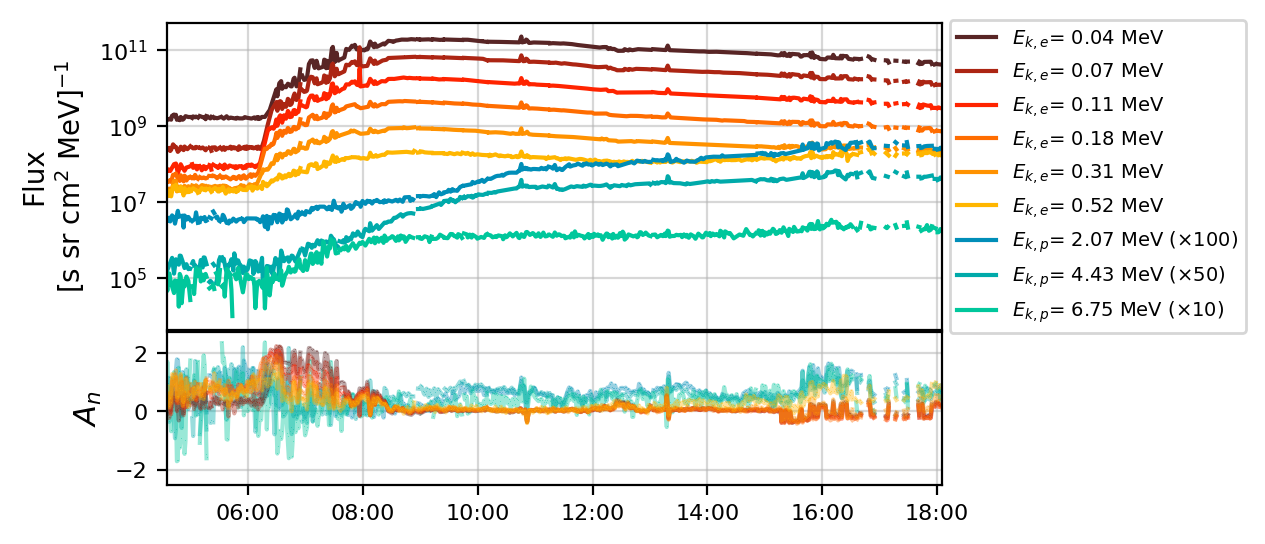}
    \caption{Electron (warm colors) and proton (cool colors) flux-time profiles of the event on 20 January 2022 from the Wind-3DP instrument with the accompanying anisotropies.}
    \label{fig:wind_eFlux_profiles}
\end{figure*}

{Figure~\ref{fig:wind_eFlux_profiles} shows the omnidirectional-intensity observations from the Wind-3DP instrument for this example event in the top panel, with the corresponding calculated anisotropy profiles in the lower panel.
The particle intensity is plotted with increasing energy channels from the top (with warm colours indicating electron measurements) to the higher proton energy channels at the bottom (in cool colors; scaled as indicated in the legend). 
The anisotropy profiles show that the values for electrons reach near zero soon after 08:00 until about 15:00. }
{The increase in anisotropy near 15:00 has been investigated by the authors and no other solar events are reported (nor found through JHelioviewer) that might cause this.}

This was chosen as the example event because the recorded Wind-3DP data shows relatively clear flux and anisotropy profiles for both electrons and protons, and because a somewhat clear image of both the flare and CME could be rendered through JHelioviewer.
There was a smaller B5.3-class flare that occurred $\sim20$~minutes before the M5.5-class flare, but due to the B-class flares falling into the `background' activity, there are no obvious impacts detected.

\section{Modelling approach and model-data comparisons} \label{sec:model}


The SEP propagation model solves a 1D focused transport equation (FTE) derived by \citet{roelof1969} to create the particle intensity time profile with corresponding anisotropy values. 
This FTE is given as
\begin{align}
    \frac{\partial f}{\partial t} = - \frac{\partial \left(\mu v f\right)}{\partial s_p} - \frac{\partial }{\partial \mu} \left( \frac{(1-\mu^2)}{2L} v f \right) + \frac{\partial}{\partial \mu} \left(D_{\mu\mu} \frac{\partial f}{\partial\mu}\right), \label{eqn:FTE}
\end{align}
where $f(t, \mu, s_p, p)$ is the distribution function, calculated over time $t$, as a function of the pitch-angle cosine $\mu$, distance along the Parker magnetic field line $s_p$, and particle momentum $p$.
The terms on the right side of Equation~\ref{eqn:FTE} are known as the streaming, focusing, and scattering terms, respectively.
The streaming term describes the fluence of particles along the Parker spiral with a particle speed $v$. 
The second term uses the magnetic focusing length $L(s_p)$ to describe the focusing of particles along a magnetic flux tube. Lastly, the scattering (or collision) term introduces the pitch-angle diffusion coefficient $D_{\mu\mu}$  \cite[see, e.g.,][and references therein]{vandenBergEA20} which is related to the pMFP of the particle using \cite[e.g.][]{BieberEA94}
\begin{align}
    \lambda_{\parallel} = \frac{3}{8}v \int_{-1} ^{+1} \frac{(1-\mu^2)^2}{D_{\mu\mu}(\mu)} \text{d}\mu. \label{eqn:MFP}
\end{align}

For $D_{\mu \mu}$, we adopt the form 
{\citep[used by][]{BeeckWibberenz1986,BieberEA94,Droge00}, }

\begin{equation}
    D_{\mu \mu} = D_0 (|\mu|^{s - 1} + H) (1 - \mu)^2,
\end{equation}

{where $s = 5/3$ is the inertial range (Kolmogorov) spectral index of the slab turbulence component (discussed in more detail later) and $H = 0.05$ is introduced to account for {dynamical turbulence or nonlinear scattering effects} leading to finite scattering through $\mu = 0$. The level of scattering, $D_0$, is obtained from $\lambda_{\parallel}$.}

The pMFP is not a variable that can be directly observed or measured and is most commonly derived theoretically or inferred through inverting SEP event observations with transport models to find the best pMFP value to fit the observed flux-time profile.


\citet{reid_1964} first noticed that during SEP events the particle diffusion is anisotropic at the onset of the event, but becomes isotropic again soon after.
This evolution of particle diffusion is seen in the lower panel of Figure~\ref{fig:wind_eFlux_profiles} where the anisotropy is high at the onset ($\sim$~06:00) but soon after ($\sim$~09:00) the anisotropy (mostly from the electron profiles) reduces to near-zero. {We will focus on this late isotropic phase of the event for the model-data comparison.} In the model, the anisotropy is calculated over time as

\begin{equation}
    A_n (s_p, t) = 3 \frac{\int_{-1}^{+1} \mu \; f(s_p,\mu,t) \; \text{d}\mu }{\int_{-1}^{+1} f(s_p,\mu,t) \; \text{d}\mu}, \label{eqn:anisotropy}
\end{equation}

where the result lies within $-3$ and $+3$ \cite[see, e.g.,][]{straussEA2022}.
This adds another restriction to the observational data as the anisotropy must approach zero soon after the event has peaked. {Unfortunately, not all instruments are able to provide time-resolved anisotropy measurements and we use a pragmatic methodology of only comparing simulation results with the observed anisotropy when it is available. The anisotropy measurements themselves are not taken into account when calculating the goodness-of-fit statistics (see the discussion below) but rather as a qualitative check that the simulation results agree reasonably with the observations.}
{Table~\ref{tab:instruments} summarizes the viewing information for each spacecraft and if corresponding anisotropy data was used in this study.}

\begin{deluxetable}{rcCCc}
\tabletypesize{\scriptsize}
\tablewidth{0pt} 
\tablecaption{ {The instrument details.}  \label{tab:instruments}}
\tablehead{
\colhead{Spacecraft} & \colhead{\multirow{2}{*}{Anisotropy}}  & \colhead{Viewing} & \colhead{Viewing} & \colhead{Rotation of}\\
\colhead{\& instrument} & \colhead{} & \colhead{directions} & \colhead{angle} & \colhead{spacecraft}
} 
\startdata 
Wind-3DP & yes & 8 & 22.5^{\circ} & yes \\
SOHO-LED$^{\dagger}$ &  no & 1 & 64^{\circ} & no \\
SOHO-HED$^{\dagger}$ &  no & 1 & 120^{\circ} & no \\
STA-SEPT & no & 4 & 52^{\circ} & no \\ 
STA-HET  & no & 1 & 55^{\circ} & no \\
\enddata
\tablecomments{ $^{\dagger}$ Information taken from \citet{TorstiEA97_sohoanis}.
}
\end{deluxetable}

{STA-SEPT, with four viewing directions, have particle telescopes directed towards and away from the Sun (labeled sun and asun, respectively) along the nominal Parker spiral, as well as north and south (perpendicular to the ecliptic plane). 
However, STA was reoriented in early 2015 such that the north and south apertures are swapped, and the sun and asun apertures now look perpendicular to the Parker spiral \citep[for more information, see][]{BadmanEA2022_STAreorientation}}.

At the specific position of an instrument measuring SEP intensity, the isotropic SEP decrease seen in the flux-time profile at later times of the event (approximately between $2-6$~hours after the event start time) is predominantly determined by the pMFP, while the radial dependence of the MFP would also significantly impact the onset of the profile \citep[][]{Duggal1979, Straussetal2017b}. The original SEP code is set up such that the radial MFP $\lambda_r$ is constant and a free parameter, whereas our work is focused specifically on the parallel MFP $\lambda_{\parallel}$ being the freely adjustable MFP parameter. For simplicity, we therefore adopt a constant value of $\lambda_{\parallel}$ in the present version of the model. 
{The modelling approach used here also neglects direct solar wind effects, both solar wind convection and adiabatic energy losses, which will influence the decay phase of the simulations for slower particles, i.e. lower energy protons, while being negligible for faster particles, i.e. higher energy protons and electrons of all energies considered here \citep[][]{Ruffolo1995}. This can in turn influence the derived pMFP values for low energy protons, maybe as much as 60\% for energies lower than 20 MeV \citep[][]{Ruffolo1995}. The magnitude of these effects depend, however, on the value of the pMFP adopted (which is varied in this work) but importantly also on the (time dependent) spectral index of the SEP distribution which is not considered in this work.}

\subsection{Free Parameters}\label{sec:freeParams}

The SEP code has several open parameters that must be adjusted for each event and particle kinetic energy range.
The parameters that need to be set for each solar event are the radial position of the observing instrument (in au), the SW speed $V_{sw}$ (we use the nominal value of 400~km s$^{-1}$), the total time to be simulated (in hours, usually $10$~hrs for electrons and $20$~hrs for protons), and the type of injection used (either a delta-function in time or a Reid-Axford temporal profile).
For each energy channel that is simulated the particle kinetic energy (in MeV) and the particle species that is being compared (either electrons or protons) must also be specified.

The injection function defines an inner boundary condition for the SEP model and specifies the time profile of the SEPs released into the inner heliosphere from the solar corona. This work uses the Reid-Axford \citep{reid_1964, axford1965} injection profile, which produces a flux-time profile that is specified at the model's inner boundary ($s_p=s_{p,0}$). 
This is defined by \citet{Droge00} as
\begin{equation}
    f_0 (s_p = s_{p,0}, t) = \frac{C}{t}\exp{\left[ - \frac{\tau_a}{t} - \frac{t}{\tau_e} \right]}, \label{eqn:reidAxford}
\end{equation}
where $\tau_a$ and $\tau_e$ are the acceleration and escape times, respectively, and $C$ is a constant that is used to normalize the simulated profile to the observations.

These time variables along with the pMFP introduced earlier make up the three free parameters that we adjust in the code to fit the simulation to the observational data, which should provide information on the conditions in the inner heliosphere during the event.
It should be kept in mind that the constant $C$ currently has no physical meaning and as such we don't draw any conclusions as to the values that are inferred for it.

\begin{figure*}[ht!]
    \centering
    \includegraphics[width=\linewidth]{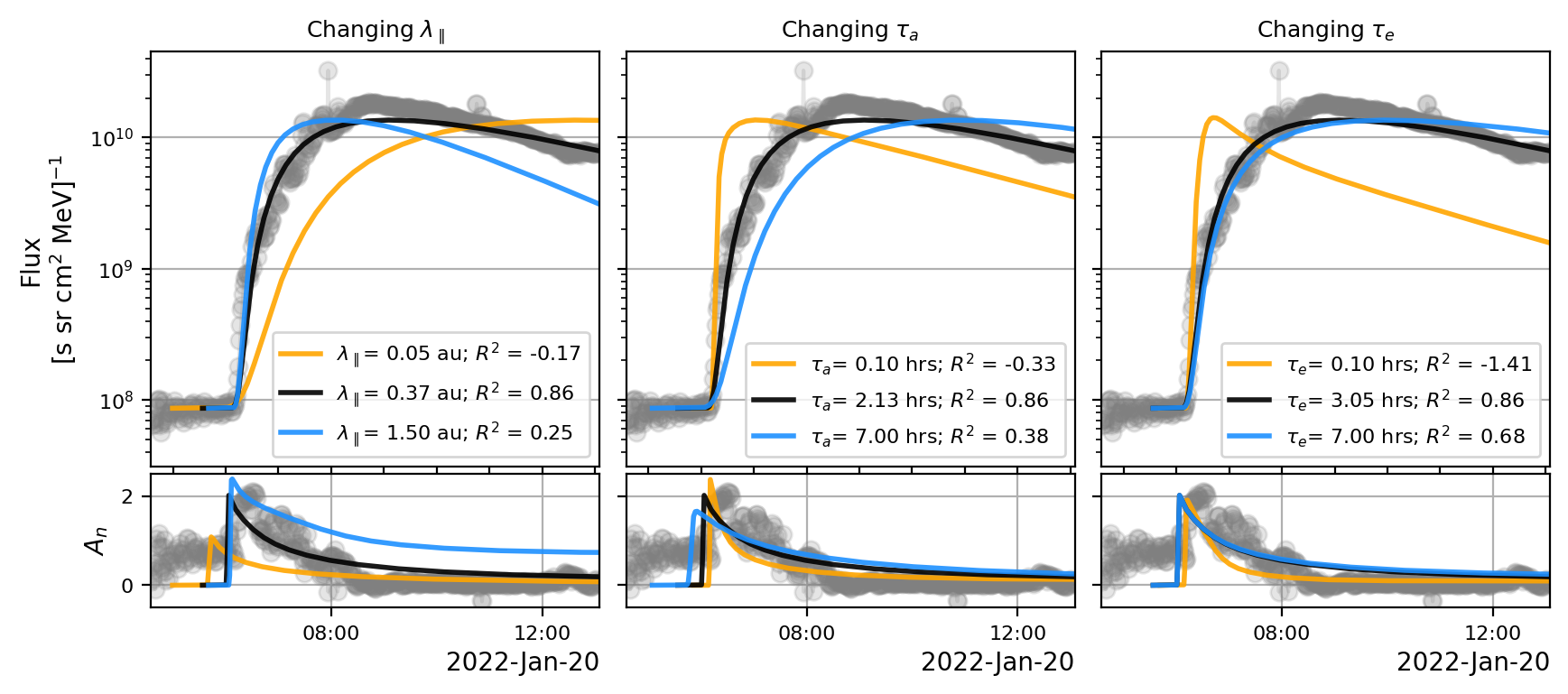}
    \caption{A guide on how each parameter impacts the flux-time profile using the Wind-3DP electron data for the $108.4$~keV energy channel recorded during the 20 January 2022 event. The best-fitting simulation is seen in black, overlaid onto the grey observations. The lower and higher values for each parameter are shown with orange and blue lines, respectively. The corresponding goodness-of-fit values are also displayed for each simulation.}
    \label{fig:parameterChanges}
\end{figure*}

The influence of each free parameter can be seen in Figure~\ref{fig:parameterChanges} which shows the $108.4$~keV Wind-3DP electron observations (grey circles) beneath the best-fitting simulation (black line).
This best-fit simulation used $\lambda_{\parallel}=0.37$~au, $\tau_a=2.13$~hrs, and $\tau_e=3.05$~hrs for the free parameters.
Each of the three parameters was decreased or increased (while keeping the other parameters to their best fitting values) to show their influence in the flux-time profile, seen with the orange and blue lines, respectively. We generally find that during the later, isotropic, phase of the SEP event, the omni-directional intensity decreases exponentially with the slope primarily determined by the level of pitch-angle scattering (i.e. the pMPF) as also confirmed by analytical approximations for an isotropic distribution \citep[][]{Duggal1979, Straussetal2017b}. {Many event and simulation-specific parameters influence the onset phase of the event. These include, for instance, the angular response of the instrument and whether the magnetic field is aligned with the instrument such that the initial anisotropic SEP beam is actually captured \citep[see e.g. the discussion by][]{brudernEA2022}, while in the model, the assumed pitch-angle dependence of $D_{\mu \mu}$, and the radial dependence of $\lambda_{\parallel}$ can also influence the initial anisotropic phase. Due to these constraints, we opt to focus on fitting the late isotropic phase of the event which we believe gives robust results for $\lambda_{\parallel}$. The values of $\tau_a$ and $\tau_e$, although derived, depend on the model set-up and details of each event and instrument and are therefore not considered robust estimations and are not considered for the rest of the study.}

{The best-fit result is often found using a \textit{by-eye method} \citep[as seen in][]{Droge00}, so to incorporate a level of accuracy, we included a more detailed goodness-of-fit test within the methodology, {as done by \citet{Ruffoloetal1998} and \citet{Bieberetal2002}, which is} discussed in more detail in the next section}.
\subsection{Goodness-of-fit}\label{sec:goodnessoffit}

{In this work, we employed the coefficient of determination goodness-of-fit (GoF) test, which is} defined by \citet{achen1982} as the fraction of explained variance over the total variance.
This hypothetical \textit{percentage of agreement} is mathematically defined by \citet{r2berryfeldman} as
\begin{align}
    R^2 &= 1 - \left(\frac{\sum_{j=1}^n \left(O_j - S_j \right)^2 }{\sum_{j=1}^n \left(O_j - \Bar{O}\right)^2} \right), \label{eqn:goodnessoffit}
\end{align}
where, in this case, the omni-directional flux observations are denoted by $O_j$ for $n$ number of observations, $S_j$ indicates the simulated (or predicted) values, and $\Bar{O}$ is the average value of the observations. 
The resulting $R^2$ value for a fit should lie within $0.0$ and $1.0$, where unity implies that the model includes all the potential impacts of the SEPs' transport and that the simulation is a perfect fit. 
Thus, an $R^2$ value closer to zero (sometimes even negative, as seen in Figure~\ref{fig:parameterChanges}) would imply that the simulation is not fitting at all.
We used a minimum threshold $R^2$ value of $0.80$ (or $80$\%) to determine if the result is a \textit{good fit} and be able to further use the inferred data (as done in Section~\ref{sec:comparisonWtheory}).

To incorporate the $R^2$ GoF into this study, we utilized the \texttt{r2\_score} function from the \texttt{scikit-learn} Python package \citep{scikit-learn}.

The GoF results when changing the free parameters are provided in Figure~\ref{fig:parameterChanges} where the best-fit returns $R^2=0.86$ (hypothetically, $86$\% of the simulated data agrees with the observed data), while the other simulations' GoF values are drastically lower; with the differences between 0.18 and 2.27.
This shows how much influence each parameter can have on the result. 
For example, $\tau_e$ increased from $3.05$~hrs to $7.00$~hrs and $R^2$ only changes by $0.18$; whereas the same change is made with $\tau_a$ and $R^2$ changes by almost $0.50$.
However, the most impactful parameter is the pMFP, where parameter changes were the smallest but in relation produced the largest change in GoF.
It is important to note that the severity of this impact also changes between events or particle energies.

\subsection{Onset determination and normalization processes} \label{sec:onsetAlignment}

\begin{figure*}[ht!]
    \centering
    \includegraphics[width=0.75\textwidth]{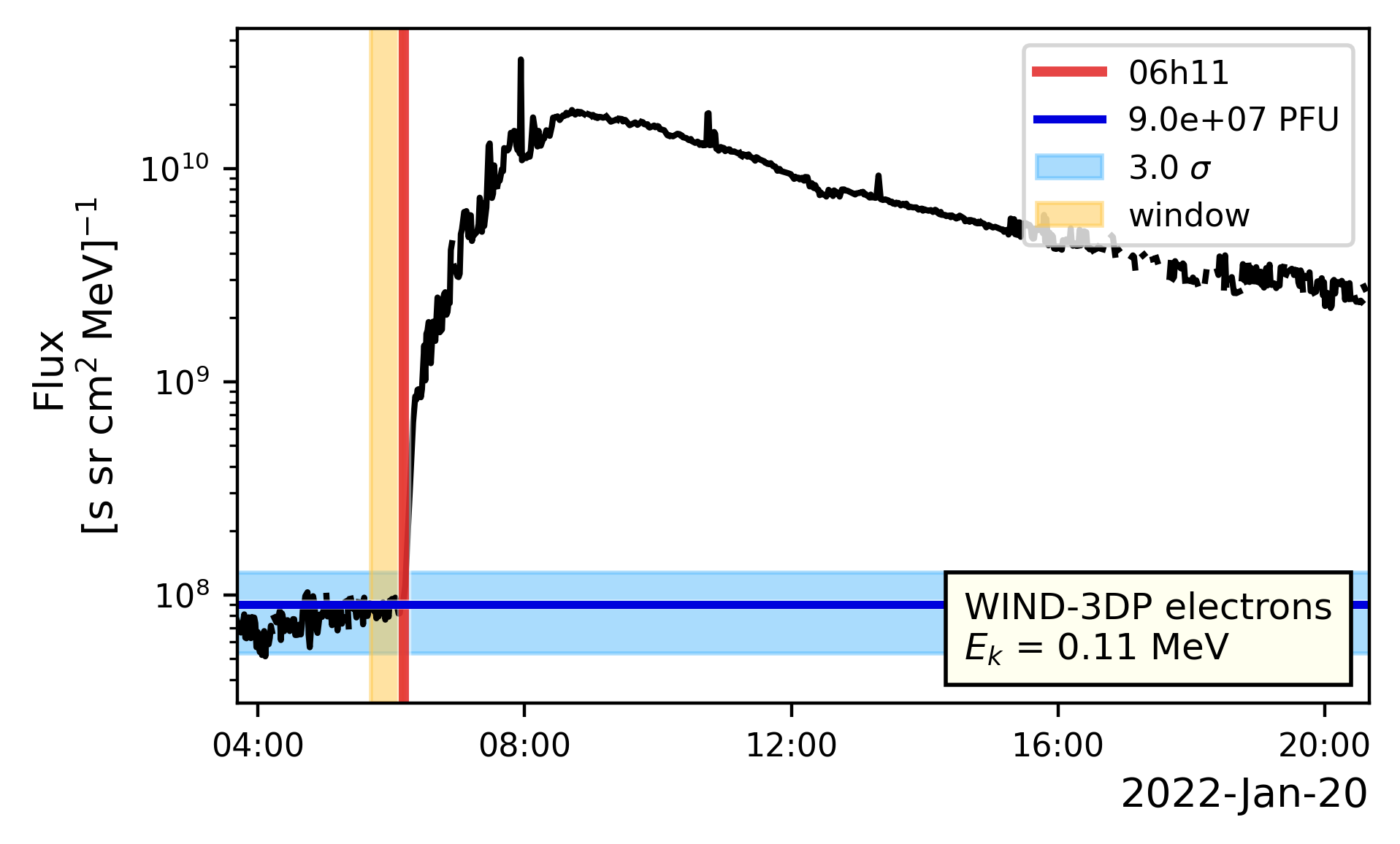}
    \caption{An illustration of how the SEP onset is determined. Additional details are given in the text.}
    \label{fig:onsetDetermination}
\end{figure*}

When an instrument starts observing SEPs at a much higher intensity than the background, it is referred to as the onset and can be seen in the time profile in the form of a steep, almost-vertical increase for the case of impulsive SEP events, as seen in Figures~\ref{fig:parameterChanges}~and~\ref{fig:onsetDetermination} at about 06:00. 
In this work, we normalized the {maximum} intensity of the model simulations and aligned the onsets within the time period, aligning the model in both the x- and y-axes. 
To normalize the particle flux we find the average flux value around the peak of the observed data and divide it by the average flux value around the peak of the simulated data to get $C$. 
The full simulation data set is then multiplied by $C$ to get the normalized flux profile. 
It is important to note that the simulation results are initially provided in arbitrary units before being multiplied by $C$ (which now has units from dividing the observed peak flux values by the model's peak flux values), after which they have particle flux units (s sr cm$^2$ MeV)$^{-1}$. 


To align the onsets of both the observation and simulation, one must first determine at what time the onsets occurred.
The onset determination process utilized in this work was adapted from the 3$\sigma$ onset determination tool presented at the SERPENTINE MSc Course in 2022\footnote{Resources from this course are available at \url{https://serpentine-h2020.eu/workshop/}}.
This tool is a basic version of the \texttt{onset\_determination}\footnote{Available at \url{https://serpentine-h2020.eu/tools/}} Python tool that utilizes the Poisson-Cumulative Sum method also designed by the SERPENTINE group \citep{palmroosEA2022_SERPENTINEdataloader}.

This 3$\sigma$ method initially requires defining a window or interval from entirely the pre-event background, for which 30 data points (i.e. 30 minutes) were used. 
From there, the average $\Bar{x}_w$ and standard deviation $\sigma_w$ of the intervals' background intensity can be calculated. The next data point $x_1$ after the window period is compared to $x_0$ where

\begin{equation}
    x_n = \Bar{x}_{w} + 3\sigma_{w} \label{eqn:onsetCalc}
\end{equation} 

for the n$^{\mathrm{th}}$ point, if $x_1 > x_0$ then the next few data points are also checked to see if the increase is steady or an outlier.
If the increase is found to be steady then the onset is determined to be at the index position of $x_1$. 
However, if $x_1 < x_0$ the window moves forward by one data point, keeping the interval 30 units long, and these steps are repeated until an onset index is returned. 
In the results presented here, the multiplication factor for $\sigma_w$ is manually adjusted for each observed data set. 
This manual adjustment is due to each data set having a different (and variable) background intensity.

{If the time profile is not defined enough then this onset determination and normalisation process will not be able to compute accurately, hence why at least one well-defined profile is necessary for the event to be considered in this study.}

Figure~\ref{fig:onsetDetermination} displays the outcome for finding the onset with the Wind-3DP electron flux profile for energy $E_k=108.4$~keV.
The vertical red line indicates where the onset was found to be, the orange region displays the final position of the window interval, the horizontal dark blue line displays what the average background intensity was determined to be before the onset (i.e. within the orange region), and the blue region shows how much of the background intensity the standard deviation considers. 
With the onset time of the observed and simulated data, one can align the two data sets (the measured and simulated SEP time profiles) and thereby infer the injection time of the SEP particles which is also the start time of the simulations, i.e. $t = 0$ in Equation~\ref{eqn:reidAxford}. 
As such, we do not need to perform a velocity dispersion analysis to infer the injection times.

\subsection{Best-Fit Results} \label{sec:bestfit_results}

\begin{figure*}[ht!]
    \centering
    \includegraphics[width=0.99\textwidth]{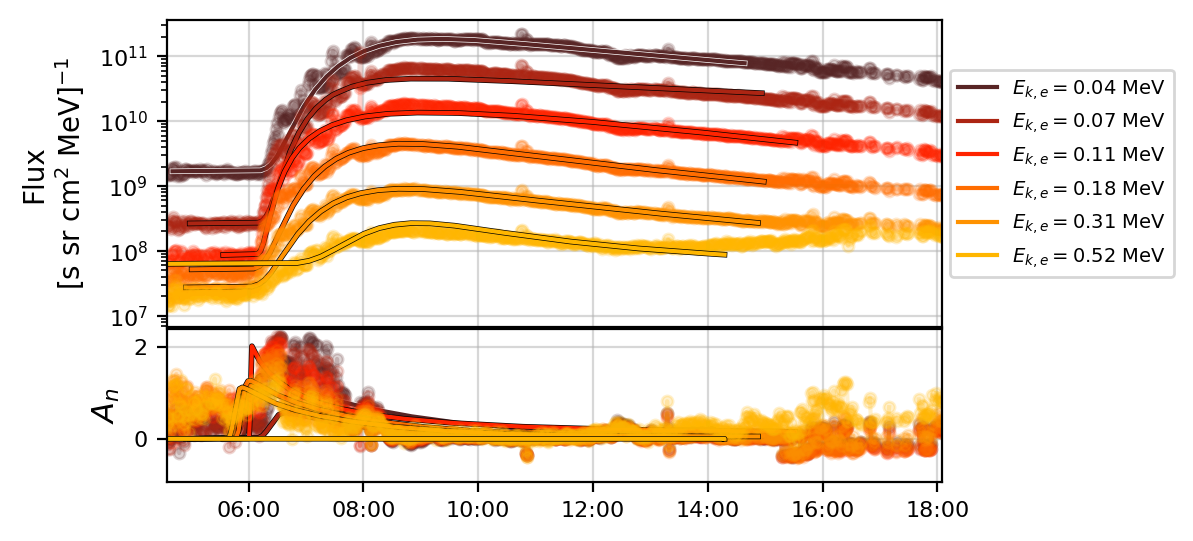}
    \caption{The results of fitting model simulations to the Wind-3DP electron data for the 20 January 2022 event. The best-fit model parameters are listed in Table~\ref{tab:example_results}.}
    \label{fig:fittedExampleEvent}
\end{figure*}

The results for fitting the Wind-3DP electron observations for the event on 20 January 2022 are seen and summarised in Figure~\ref{fig:fittedExampleEvent} and Table~\ref{tab:example_results}, respectively. The best result for this event's fit is seen with the lowest energy bin ($E_{k,e}=0.04$~MeV, the top-most intensity in the darkest red in Figure~\ref{fig:fittedExampleEvent}) where the GoF value reaches $97$\%.
While the worst result is the highest energy bin ($E_{k,e}=0.52$~MeV, the bottom-most intensity in yellow in Figure~\ref{fig:fittedExampleEvent}), where the start time is determined to be almost $1.5$~hrs earlier than the recorded event time, yet the model onset is clearly much later.
This late onset time is most likely due to the initial slope of the particle flux profile not being steep enough to be accurately determined.


The GoF method defined in Section~\ref{sec:goodnessoffit} provides most of its focus in the tail end of the time profile (i.e. from the peak to $\sim 6$~hrs later) due to the range of fluxes compared in a log-scale.
This later, isotropic phase is where the pMFP values are most defined, as seen in Figure~\ref{fig:parameterChanges} where there are smaller changes to the parameter value but the slope of the tail is shifted from a near horizontal (orange) line (where $\lambda_{\parallel}=0.05$~au) to the almost $45^{\circ}$ diagonal (blue) line (where $\lambda_{\parallel}=1.50$~au). That being said, it is worthwhile to note that the highest energy bin experiences a small steady increase from about 13:00 that the simulated profile is trying to fit, which is the main cause for the low $R^2$ score.

\begin{deluxetable}{RCCCCc}
\tabletypesize{\scriptsize}
\tablewidth{0pt} 
\tablecaption{The results for fitting the Wind-3DP electron intensities in various energy bins for the 20 January 2022 event.  \label{tab:example_results}}
\tablehead{
\colhead{Energy} & \colhead{$\lambda_{\parallel}$} & \colhead{$\tau_a$} & \colhead{$\tau_e$}  & \colhead{\multirow{2}{*}{$R^2$}} & \colhead{Start}\\ 
\colhead{[MeV]} & \colhead{[au]} & \colhead{[hrs]} & \colhead{[hrs]} &  & \colhead{time}
} 
\startdata 
0.04 & 0.28 & 13.24 & 0.64 & 0.97 & 04:40 \\ 
0.07 & 0.08 &  9.80 & 0.15 & 0.79 & 04:58 \\ 
0.11 & 0.37 &  2.13 & 3.05 & 0.86 & 05:33 \\ 
0.18 & 0.23 &  9.96 & 0.60 & 0.96 & 05:00 \\ 
0.31 & 0.18 &  9.98 & 0.60 & 0.96 & 04:54 \\ 
0.52 & 0.36 & 51.19 & 0.28 & 0.65 & 04:19    
\enddata
\tablecomments{Note that the recorded x-ray onset (implying the event eruption time) is 05:41. }

\end{deluxetable}


Referring to just the pMFP values that correspond to acceptable GoF values ($R^2\geq 0.80$), we can notice a similar trend to that of \citet{Droge00} where the electron pMFP values are steadily decreasing in value with an increase in energy, despite the outlier.

\subsection{Summary of pMFP Results} \label{sec:mfp_results}

{Over $200$ observational data sets were processed in this study from $15$ events, using data from three spacecraft and five instruments.
While the full energy range for each instrument (and species) was considered, not all of the energy channels produced results with acceptable GoF values.
Table~\ref{tab:data_channels} summarizes the energy channels from the respective instruments that provided acceptable GoF values. The arithmetic mean for each energy range was used for the simulations in the SEP transport model.}

\begin{deluxetable*}{||c||cc||cc||cc||}
\tabletypesize{\scriptsize}
\tablewidth{0pt} 
\tablecaption{ {The energy channels of the instruments used.}  \label{tab:data_channels}}
\tablehead{
\colhead{Wind-3DP} & \multicolumn{2}{c}{SOHO p$^+$} & \multicolumn{4}{c}{STEREO-A} \\
\colhead{\underline{energy ranges}} & \multicolumn{2}{c}{\underline{energy ranges}} & \multicolumn{4}{c}{\underline{energy ranges}} \\
\colhead{e$^-$} & \colhead{LED} & \colhead{HED} & \colhead{HET e$^-$} & \colhead{HET p$^+$} & \colhead{SEPT e$^-$} & \colhead{SEPT p$^+$}
} 
\startdata 
0.02-0.04 & 1.3-1.6   & 13-16   & 0.7-1.4 & 13.6-15.1 & 0.045-0.055 & 0.220-0.246 \\
0.03-0.05 & 2.0-2.5   & 16-20   & 1.4-2.8 & 15.0-17.1 & 0.055-0.065 & 0.246-0.273 \\
0.05-0.09 & 3.2-4.0   & 20-25   & 2.8-4.0 & 17.0-19.3 & 0.065-0.075 & 0.273-0.312 \\
0.08-0.14 & 4.0-5.0   & 25-32   & \nodata & 20.8-23.8 & 0.075-0.085 & 0.312-0.351 \\
0.13-0.23 & 5.0-6.4   & 32-40   & \nodata & 23.8-26.4 & 0.085-0.105 & 0.351-0.390 \\
0.21-0.40 & 6.4-8.0   & 40-50   & \nodata & 26.3-29.7 & 0.105-0.125 & 0.390-0.438 \\
0.36-0.66 & 8.0-10.0  & 50-64   & \nodata & 29.6-33.4 & 0.125-0.145 & 0.438-0.496 \\
\nodata   & 10.0-13.0 & 64-80   & \nodata & 33.4-35.8 & 0.145-0.165 & 1.767-1.985 \\
\nodata   & \nodata   & 80-100  & \nodata & 35.6-40.5 & 0.165-0.195 & 1.985-2.224 \\
\nodata   & \nodata   & 100-130 & \nodata & 40.0-60.0 & 0.195-0.225 & 2.224-6.500 \\
\nodata   & \nodata   & \nodata & \nodata & 60.0-100.0& 0.225-0.255 & \nodata \\
\nodata   & \nodata   & \nodata & \nodata & \nodata   & 0.255-0.295 & \nodata \\
\nodata   & \nodata   & \nodata & \nodata & \nodata   & 0.295-0.335 & \nodata \\
\nodata   & \nodata   & \nodata & \nodata & \nodata   & 0.335-0.375 & \nodata \\
\nodata   & \nodata   & \nodata & \nodata & \nodata   & 0.375-0.425 & \nodata \\
\enddata
\tablecomments{
All energies are provided in MeV.
}

\end{deluxetable*}

\begin{figure*}[ht!]
    \centering
    \includegraphics[width=\linewidth]{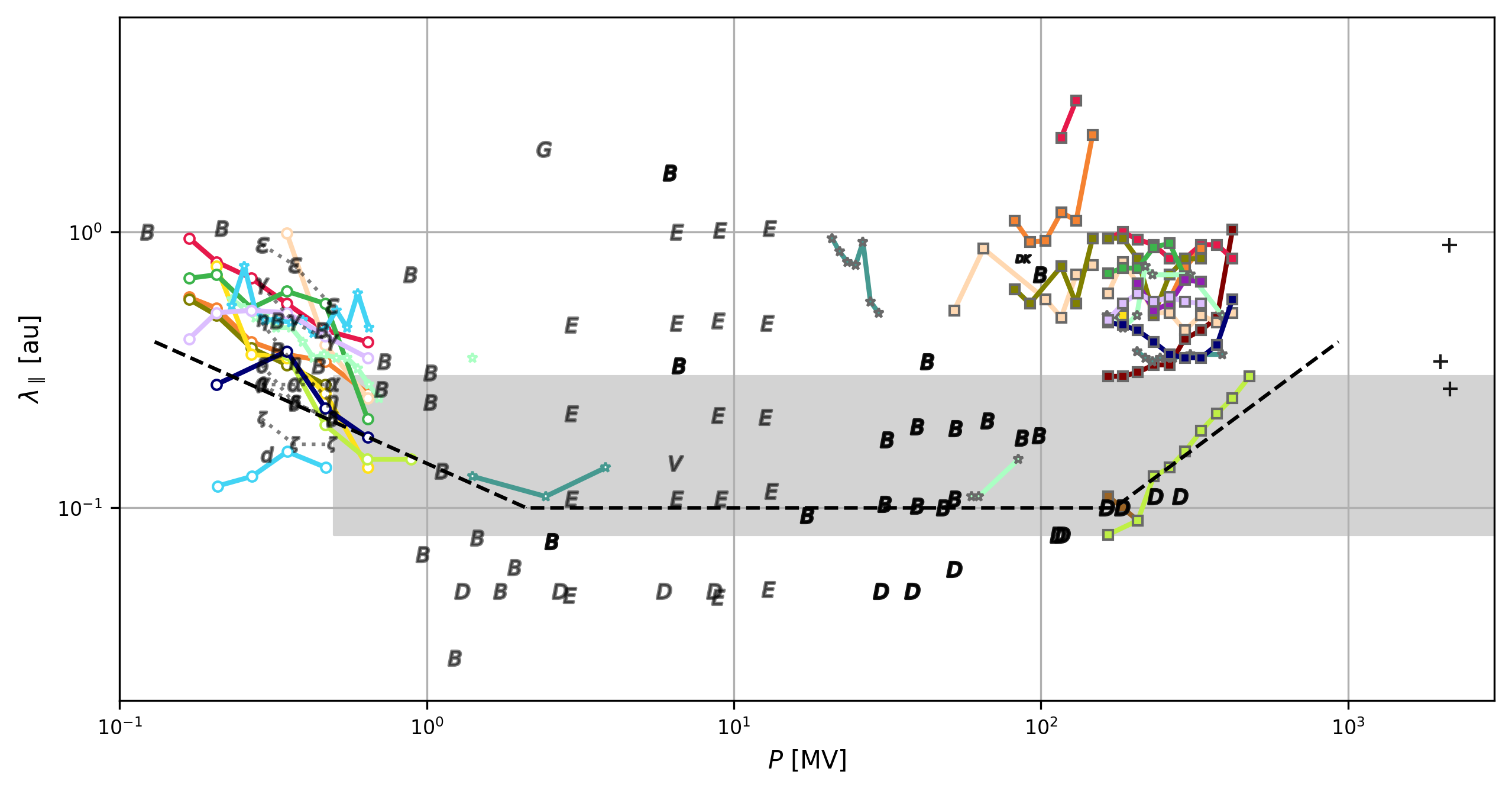}
    \caption{Summary of all the pMFP results from fitting the events listed in Table~\ref{tab:events_fitted} as a function of their corresponding rigidity (in MV). The particle species are predominantly separated into the left and right sides of the figure; the electrons are open markers on the left and the protons are closed-face markers on the right. Each color indicates an event (corresponding to the colors in Figure~\ref{fig:solarCycle}), and each marker style indicates a different spacecraft; SOHO is denoted by square markers, Wind by circles, and STA by stars. The dashed black line marking a valley shape provides the trend seen by \citet{Droge00}. The grey-shaded region shows the so-called Palmer consensus values \citep[][]{Palmer82}. See text for external data point details. }
    \label{fig:mfpVrig}
\end{figure*}

A summary of the best-fitting pMFP results is shown in Figure~\ref{fig:mfpVrig}, where the electron values are found with open markers at lower rigidities (to the left), and the proton results are found with closed-faced markers at higher rigidities (to the right). 
Results from previous studies which fitted various observations have been included for the purposes of comparison with the results presented here. 
The dashed line represents the pattern reported by \citet{Droge00}, where with increasing rigidity the electron pMFPs decrease to a lower boundary and proton pMFPs increase from that boundary.
Letter markers (B, G, D, DK, d, V, and E) indicate observations from \citet{BieberEA94}, \citet{GloecklerEA95}, \citet{Droge00}, \citet{DrogeKartavykh2009}, \citet{drogeEA14}, \citet{Vogtetal2020}, and \citet{Engelbrecht_2022}, respectively, while Greek letter markers ($\alpha, \beta, \gamma, \delta, \epsilon, \zeta, \eta$) indicate data points taken from \citet{AguedaEA2014}\footnote{Using $\lambda_r = \lambda_{\parallel} \cos^2\psi$ to convert from radial to parallel MFP, where we assume a nominal SW speed of $V_{sw}=400$~km s$^{-1}$, resulting in $\psi=45^{\circ}$.} in temporal order, starting with $\alpha$ for the oldest event. 
{Bold font markers for the external data points are indicative of proton pMFP values, except in cases not referring to SEP results, these include pick-up ions from \citet{GloecklerEA95}, and Jovian electrons from \citet{Vogtetal2020} and \citet{Engelbrecht_2022}.}
Results from fitting GLE events from \citet{Bieberetal2002}, \citet{BieberEA2004}, and \citet{SaizEA2005} are included as $+$ markers, as well as the consensus range defined by \citet{Palmer82} as the grey shaded region.

\begin{figure*}[ht!]
    \centering
    \includegraphics[width=0.46\linewidth]{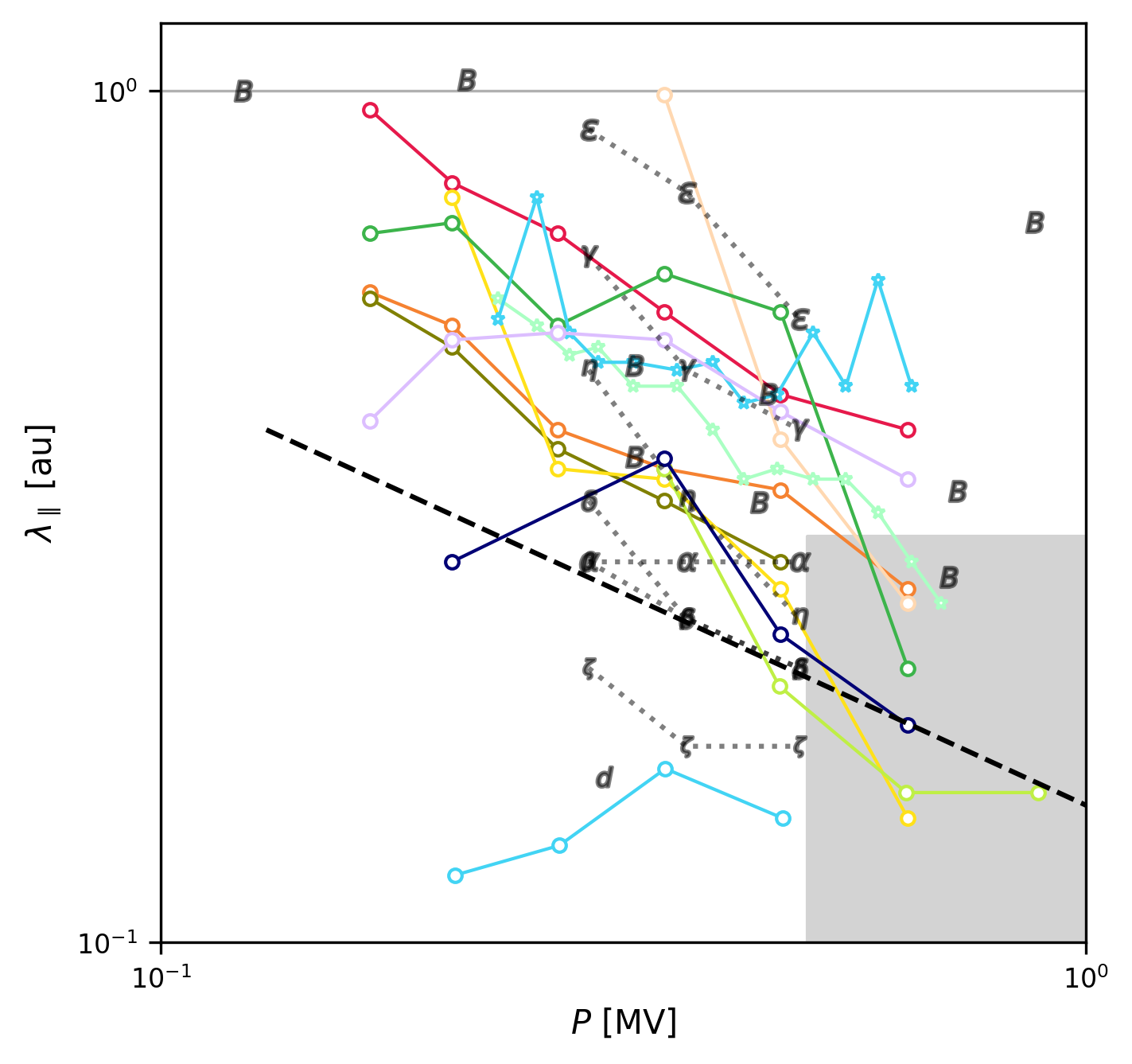}
    \includegraphics[width=0.45\linewidth]{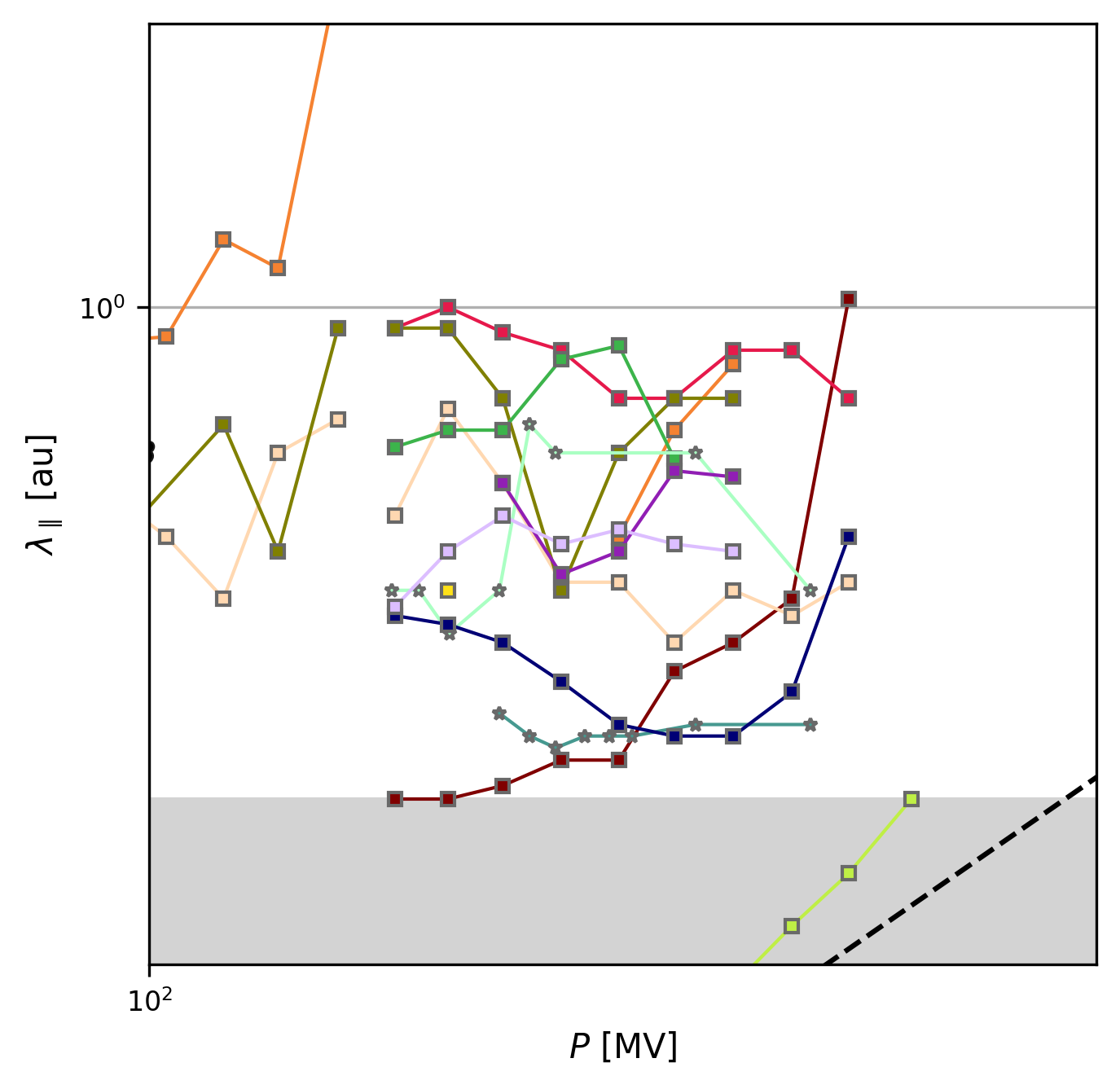}
    \caption{A zoomed-in view of the left and right of Figure~\ref{fig:mfpVrig}. All labels and references remain the same.}
    \label{fig:mfpVrig_zoom}
\end{figure*}

Each color indicates a different event fitted in this work, while each line is indicative of a single instrument's observation of a species for one event over a range of rigidities. 
For example, the bright green squares that are nearly isolated on the right of Figure~\ref{fig:mfpVrig} show the SOHO/HED proton pMFPs for the event on 17 May 2012, while the bright green circles on the left (better seen in the left panel of Figure~\ref{fig:mfpVrig_zoom}) show the Wind-3DP electron pMFPs for the same event.
A basic reference for when each event occurs with regards to the color can be found using Figure~\ref{fig:solarCycle}.

Converting the corresponding energies to rigidities, forces the Larmor radius to be the same for both electrons and protons for a given rigidity. With that consistency, we can compare both species in the same figure.
Rigidity $P$ is calculated as $P = \sqrt{E_k(E_k + 2E_0)}$, where $E_k$ is the particle's kinetic energy (in units of eV) and $E_0$ is the rest mass energy \citep{moraal2013}.
The ranges of kinetic energies covered in Figure~\ref{fig:mfpVrig} are $E_{k,e}=0.02-4.0$ MeV for electrons (the corresponding rigidity range is calculated as $P_e=0.14-4.48$ MV) and $E_{k,p}=1.30-130$ MeV for protons ($P_p=49.4-510.7$ MV).

For both particle species considered, the results of prior studies, as well as those of the present study, display considerable inter-event variations, ranging over approximately a full order of magnitude. For most of the proton pMFPs no clear rigidity dependence can be discerned, although those from the SOHO/HED proton event on 17 May 2012 display a clear dependence in agreement with the trend observed by \citet{Droge00}. Proton pMFPs reported on here remain mostly well above the Palmer consensus range. For electrons at the lowest rigidities, pMFPs display an increase with decreasing rigidity, in agreement with the trend reported in prior studies.

\section{Comparison with Theory} \label{sec:comparisonWtheory}

\begin{figure*}[ht!]
    \centering
    \includegraphics[width=0.48\linewidth]{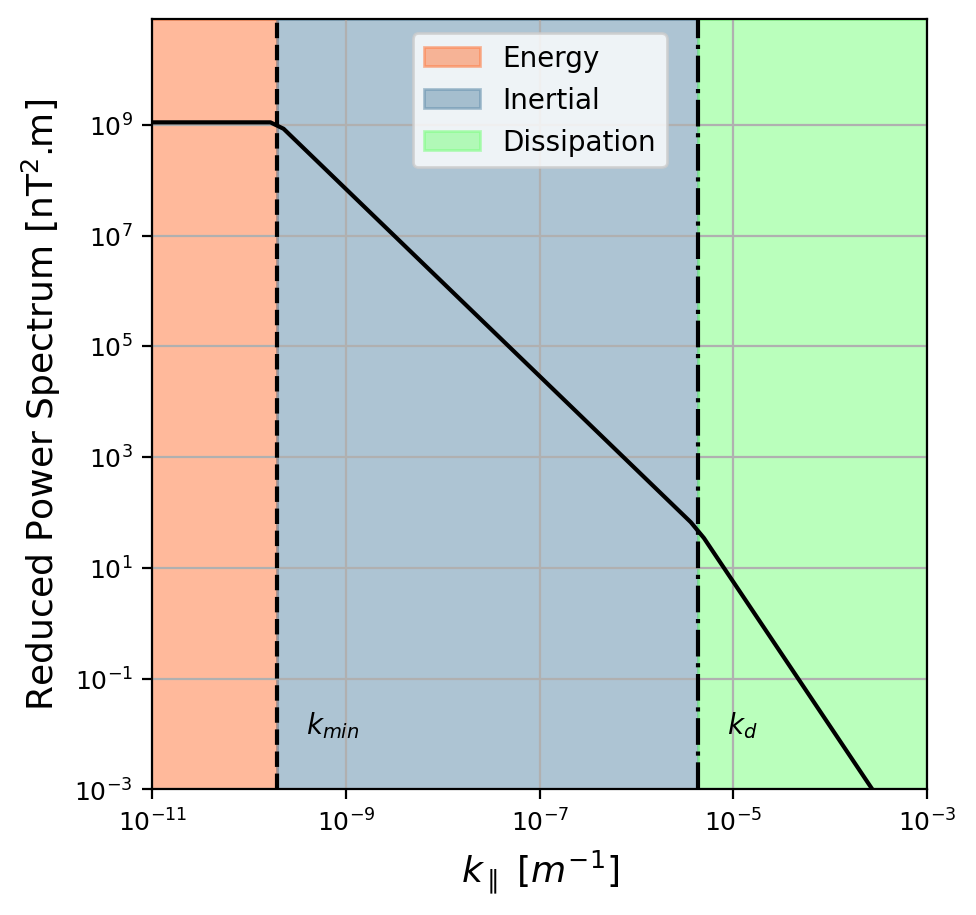}
    \includegraphics[width=0.48\linewidth]{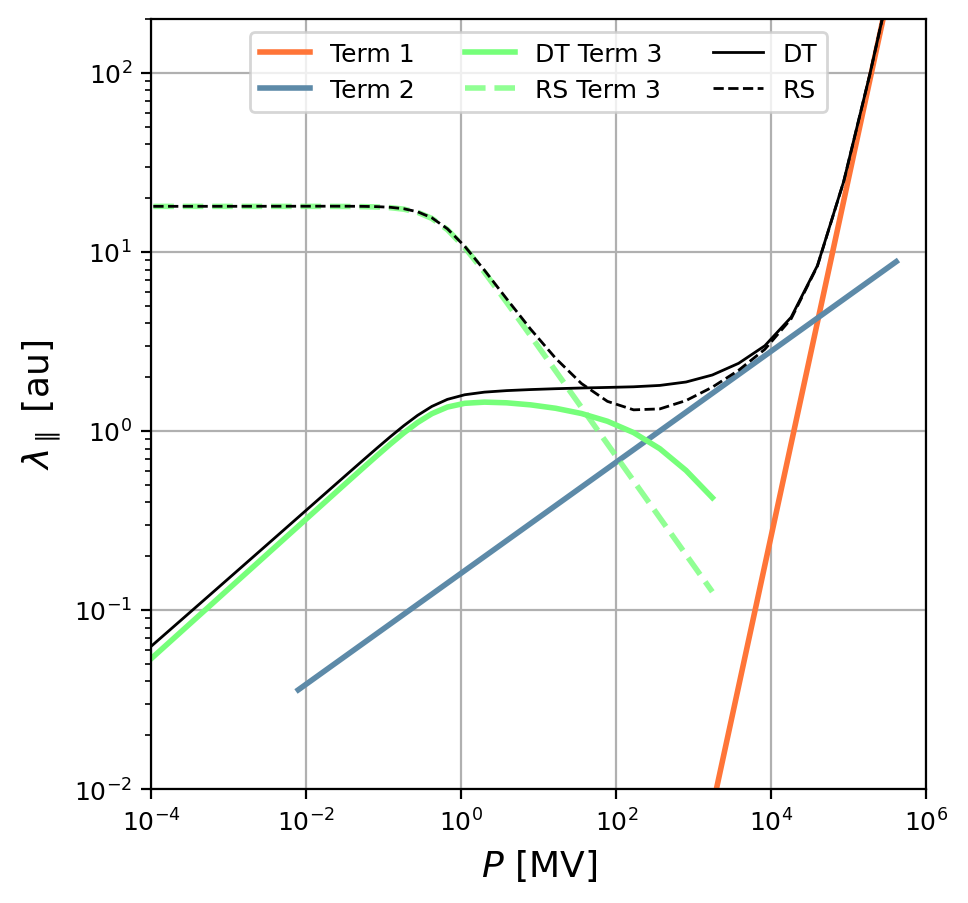}
    \caption{The left panel shows the reduced power spectrum as given in \citet{EngelbrechtBurger13} and plotted using Equations~\ref{eqn:reduced_power_spectrum} and \ref{eqn:g0}. The right panel shows which term within the square brackets of Equation~\ref{eqn:turb_mfp} corresponds to which region in the power spectrum by use of coordinating colors. See text for details.}
    \label{fig:reduced_power_spectrum}
\end{figure*}

To compare the pMFP values with expected theoretical predictions, as well as ascertain the degree to which variations due to the event-to-event variations of, e.g., turbulence quantities translate to variations in electron transport coefficients, we employ two expressions for electron pMFPs based on those derived from the Quasilinear Theory \cite[QLT,][]{Jokipii66} by \citet{TS02,TS03}, motivated partly by the fact that these expressions, at least qualitatively, are in agreement with previous model-based estimates \cite[see, e.g.,][]{BieberEA94,Droge00,Engelbrecht_2022}, as well as having been employed in electron modulation studies \cite[e.g.][]{EngelbrechtBurger13,Engelbrecht19}. 
These expressions are derived assuming an observationally-motivated \cite[see, e.g.,][]{BrunoCarbone16} form for the slab turbulence power spectrum with a wavenumber-independent energy-containing range, and inertial range with spectral index $s$, and a dissipation range with spectral index $p$.
This is illustrated in the left panel of Figure~\ref{fig:reduced_power_spectrum} and given by \citet{TS03} as
\begin{eqnarray}
    g_{slab}(k_{\parallel}) &= 
    \begin{cases}
        g_0 \; k_{min}^{-s}, &\text{for } |k_{\parallel}| \leq k_{min} ;\\
        g_0 \; |k_{\parallel}|^{-s}, &\text{for } k_{min} \leq |k_{\parallel}| \leq k_d;\\
        g_0 \;k_d^{p-s} \; |k_{\parallel}|^{-p}, &\text{for } |k_{\parallel}| \geq k_d,
        \end{cases} \label{eqn:reduced_power_spectrum}
\end{eqnarray}
where
\begin{align}
    g_0 &= \frac{\delta B_{slab}^2 \; k_{min}^{s-1} \; (s-1)}{8 \pi}\left[s + \frac{s-p}{p-1}\left(\frac{k_{min}}{k_d}\right)^{s-1} \right]^{-1} .\label{eqn:g0}
\end{align}
Note that, in the above, $\delta B_{slab}^2$ denotes the (slab) magnetic variance \cite[as opposed to 2D; see][]{MattEA90,BieberEA96}, while $k_{min}$ and $k_d$ respectively denote the inertial and dissipation range onset wavenumbers. 

Furthermore, both expressions are derived using different models for dynamical turbulence, viz. the damping and random sweeping models \cite[see, e.g.,][]{BieberEA94,ShalchiEA04dyn,DempersEngelbrecht20}. 
Essentially, the dynamical turbulence model employed specifies the time dependence of the resonance function used to calculate the pitch-angle diffusion coefficient by means of imposing a temporal decay on the turbulence correlation function, and hence, via Equation~\ref{eqn:MFP}, the pMFP \citep{BieberEA94,TS02}. 
For the random sweeping (RS) model, the time dependence of the turbulence correlation function is modeled as $\exp{\left[-\alpha_D^{2} \; k^{2}_{\parallel} \;V^{2}_{A} \;t^{2}\right]}$, while for the damping (DT) model, it is $\exp{\left[-\alpha_D\; |k_{\parallel}| \;V_{A}\;|t|\;\right]}$, with $\alpha_D \in [0,1]$ an essentially free parameter adjusting the strength of the dynamical effects \citep{BieberEA94}, and $V_A$ the Alfv\'en speed. 
In this work, we varied $\alpha_D$ to be (0.1, 0.5, 1.0) to observe if there were any drastic changes to the profile when adjusting the strength of dynamical effects.
These assumptions lead to electron pMFP expressions that differ considerably at low energies \cite[see][]{EngelbrechtEA22b}. In various limits, the results of \citet{TS03} can be employed to construct expressions for electron (and proton) pMFPs \cite[see, e.g.,][]{BurgerEA08,EngelbrechtBurger13}, given by
\begin{eqnarray}
    \lambda_{\parallel} &=& \frac{3 \; s \; R_L^2 \; k_{min}}{4 \; \pi(s-1)}   \Bigg( \frac{B_0}{\delta B_{slab}} \Bigg)^2 \nonumber \\
    & \times & \Bigg[ 1  + \frac{8}{(2-s)(4-s)}\frac{1}{R^s} + \frac{4 \; \mathcal{K}_{\mathrm{RS/DT}}}{Q^{p-s}R^s} \Bigg] ; \label{eqn:turb_mfp} \\
    \mathcal{K}_{\mathrm{RS}} &=&  \Bigg( \frac{ \sqrt{\pi}}{\; \Gamma(p/2)} + \frac{1 }{(p-2)} \Bigg)  \left( \frac{a}{2} \right)^{p-2} , \label{eqn:turb_mfp_rs}\\
    \mathcal{K}_{\mathrm{DT}} &=& _2F_1\left(1; \frac{1}{p-1}; \frac{p}{p-1}; -\frac{a}{f_1 \; Q}\right)  \frac{ a}{ f_1 }  , \label{eqn:turb_mfp_dt}
\end{eqnarray}
where subscripts `RS' and `DT' denote the applicable dynamical turbulence model, with the appropriate value of $\mathcal{K}$. 
Furthermore, the following relations are also required to complete the pMFP calculations:
\begin{eqnarray}
    f_1 = \frac{2 (p - s)}{\pi (p - 2)(2 - s)} &;& \;\;\;\;\;\;\;
    a = \frac{v}{\alpha_D \; V_A}; \nonumber \\
    R = R_L\;k_{min} &;& \;\;\;\;\;\;\; 
    Q = R_L\;k_d; \nonumber \\
    v = \frac{P\;c}{E_k + E_0} &;& \;\;\;\;\;\;\;
    R_L = \frac{P}{B_0\; c}, \nonumber
\end{eqnarray}
where $R_L$ denotes the maximal particle Larmor radius, $c$ the speed of light, and $B_0$ the background magnetic field strength. 
Expressions for proton pMFPs can be acquired from the above by setting $\mathcal{K}=0$. 
Dynamical effects are not expected to influence the parallel diffusion of protons, given their larger Larmor radii, hence proton pMFPs for both dynamical turbulence models are the same. 
It can be seen that these expressions for electrons, are comprised of three terms, representing three rigidity dependencies, and shown as a function of rigidity in the right panel of Figure~\ref{fig:reduced_power_spectrum}{.} 
Term 1, displaying a $\sim P^2$ dependence often reported on in prior studies \cite[see, e.g.,][]{Palmer82}, represents the pMFP of particles being resonantly scattered by the energy-containing range of the assumed power spectrum (orange band in left panel of Figure~\ref{fig:reduced_power_spectrum}); Term 2, with a $\sim P^{2-s}$ dependence, corresponds to the pMFP of those particles resonantly scattered by inertial range fluctuations (blue band in the left panel of the figure); while Term 3 represents the pMFP of particles resonantly scattered by dissipation range fluctuations (green band on left panel of the figure), displaying a flat rigidity dependence, and depending on which model of dynamical turbulence is assumed. 
Note that for protons only the first two terms are applicable. 

These expressions are plotted using the nominal values for various turbulence quantities discussed below in Section~\ref{sec:inputParameters} and tabulated in Table~\ref{tab:1au_params}. Due to being constructed from the original results from \citet{TS03}, they are somewhat larger than the numerical points. 
At high and intermediate rigidities, for both dynamical turbulence models and particle species, the pMFPs scale as $\sim P^2$ and $\sim P^{1/3}$, respectively, as expected from magnetostatic QLT \cite[see, e.g.,][]{ShalchiEA04} assuming a Kolmogorov inertial range spectral index. 
At the lowest rigidities shown, both pMFPs become essentially independent of rigidity, with the use of the RS model leading to pMFPs considerably larger than those yielded by the DT model of dynamical turbulence.

\subsection{Input parameters}\label{sec:inputParameters}

As the pMFP expressions given above depend on various turbulence quantities known to vary considerably at $1$~au, not only due to an inherent solar cycle dependence \cite[e.g.][]{ZhaoEA18,EW20,BurgerEA22}, but also on smaller temporal and spatial scales \cite[e.g.][]{IsaacsEA15,OughtonEA15,CuestaEA22}, a measure of the possible variability of the pMFP can be computed by taking into account said variations in turbulence as well as larger scale plasma quantities such as the HMF magnitude. 
Therefore, an ensemble of such values is here constructed based on spacecraft observations reported on by several studies, to serve as inputs for the pMFP expressions discussed above, thereby providing a measure of the potential variability of these pMFPs. 
An average value for a particular quantity is calculated, and a range of variation is computed from the standard deviations of the distributions of the observed, such that we employ a maximum and minimum value that corresponds to either a $1\sigma$ or $2\sigma$ deviation on that distribution. 
These values, tabulated in Table~\ref{tab:1au_params}, are used as inputs for Equations~\ref{eqn:turb_mfp}~-~\ref{eqn:turb_mfp_dt} for electrons and protons, and are discussed in more detail below. 

For observational inputs as to the magnetic field magnitude $B_0$ and (half) the total transverse magnetic variance $\delta B^2_N$ (computed from the N-component of the observed HMF), we employ the yearly-averaged values for these quantities reported for the period spanning $1974-2020$ by \citet{BurgerEA22}. 
To calculate the slab variance $\delta B^2_{slab}$ appropriate to Equations~\ref{eqn:turb_mfp}~-~\ref{eqn:turb_mfp_dt}, an assumption needs to be made as to the spectral anisotropy. 
We assume a 20/80 ratio for slab/2D variances, following \citet{BieberEA94}, which falls well within the range of values reported for this ratio by \citet{OughtonEA15} for the intermediate range of SW speeds applicable to this study. Then, we use a simple relationship $\delta B_{slab}^2 = 0.2\; \delta B^2$.

In terms of turbulence power spectral indices, for the inertial range spectral index $s$ we employ the value reported by \citet{BurgerEA22}, using the error bars to calculate upper and lower values for this quantity's range, while for the dissipation range spectral index $p$ we follow the same approach with the observational values reported by \citet{SmithEA2006}. 
However, for the hyper-geometric function seen in Equation~\ref{eqn:turb_mfp_dt} to converge $p$ must be greater than 2, thus the range chosen for this quantity becomes {$p=2.61^{+0.96}_{-0.60}$}. 

Several assumptions also need to be made to calculate the inertial range onset wavenumber $k_{min}$. 
To that end, we employ the correlation scale observations reported by \citet{WicksEA2010}. 
These observations numerically correspond closely to the values for the 2D correlation scale found by \citet{WeygandEA2011} in their analysis of multi-spacecraft observations at $1$~au, and thus will be assumed to correspond to this quantity in the present study \cite[see also][]{EW20}. 
The slab correlation scale is then estimated assuming the relationship between these quantities reported by \citet{WeygandEA2011}, such that $\lambda_{slab}=(2.55\pm0.76)\lambda_{2D}$. 
The inertial range onset wavenumber can then be calculated, assuming that $k_{min} \ll k_d$, from the expression for the correlation length of the three-part slab spectrum used in this work, i.e. \citet{engelbrechtPhD},
\begin{equation}
    \lambda_{slab} = \frac{\pi (s-1)}{2 k_{min}} \left[ s + \frac{s-p}{p-1} \left( \frac{k_{min}}{k_d} \right)^{s-1} \right]^{-1}
\end{equation}
which then reduces to $k_{min}=\pi (s-1)/2s\lambda_{slab}$. 
For the dissipation range onset wavenumber $k_d$, values for the dissipation range onset length-scale reported by \citet{SmithEA2011} are employed using $k_{D}=2\pi\lambda_{D}^{-1}$. 
Lastly, Equations~\ref{eqn:turb_mfp}~-~\ref{eqn:turb_mfp_dt} require as input the {Alfv\'en} speed. 
This is calculated from yearly averaged Advanced Composition Explorer 
\cite[ACE, ][]{StoneEA98} observations of the SW particle density and HMF magnitude. Note that the results for $B_0$, $\delta B_{slab}^2$ and $V_A$ from \citet{wrenchEA24} lie well within our calculated ranges. {It should also be noted that the parameter ranges for the magnetic variances and correlation scales listed in Table~\ref{tab:1au_params} include the solar cycle-related variations in these quantities reported on by several studies \cite[see][]{ZhaoEA18,EW20,BurgerEA22}.}

\begin{deluxetable*}{rCCc}
\tabletypesize{\scriptsize}
\tablewidth{0.5\columnwidth} 
\tablecaption{The ranges of observed $1$~au parameter values employed as inputs for the pMFP expressions considered in this study.  \label{tab:1au_params}}
\tablehead{
\colhead{\multirow{2}{*}{Variable name}} & \colhead{\multirow{2}{*}{Symbol}} & \colhead{Range}  & \colhead{\multirow{2}{*}{Units}} \\
 & & \colhead{considered} & 
} 
\startdata 
Background magnetic field                & B_0                              & 6.27 \pm 2.30     & nT  \\
Variance of normal magnetic field   & \delta B_N^2          & 7.95 \pm 5.37     & nT$^2$  \\
Variance of total magnetic field        & \delta B^2               & 15.91 \pm 10.74 & nT$^2$  \\
Variance of slab magnetic field         & \delta B_{slab}^2  & 3.18 \pm 2.15    & nT$^2$  \\
Inertial range spectral index $^{\dagger}$          & s                                   & 1.69 \pm 0.04   & --  \\
Dissipation range spectral index $^{\dagger}$ & p                                     &  {$2.61^{+0.96}_{-0.60}$}         & --  \\
Dissipation scale {wavenumber}        & k_d                               & 4.28 \pm 4.16    & $\times10^{-3}$~km$^{-1}$  \\
{2D correlation scale}       & \lambda_{2D}          & 1.10 \pm 0.49     & $\times10^{6}$~km  \\
Slab correlation scale                          & \lambda_{slab}        & 2.81 \pm 1.51     & $\times10^{6}$~km  \\
Inertial range onset {wavenumber}   & k_{min}                   & 2.28 \pm 1.22     & $\times10^{-7}$~km$^{-1}$  \\
Particle density                                    & \rho                          & 5.95 \pm 1.79     & $\times 10^{15}$~km$^{-3}$  \\
Alfv\'{e}n speed                                & V_A                           & 56.77 \pm 22.59    & km s$^{-1}$  \\
\enddata
\tablecomments{\\$^{\dagger}$ Calculated using a 1$\sigma$ deviation for the errors/range considered as reported by sources, whereas the other distributions used a $2\sigma$ deviation.}


\end{deluxetable*}

\subsection{Energy dependence} \label{sec:energyDependence}

All possible combinations of the parameters discussed above, including the upper and lower bounds of the uncertainties, are systematically employed as inputs for the pMFP expressions considered here. It should, however, be noted that the abovementioned parameters are often interrelated. For example, an increase in magnetic variance is often associated with an increase in HMF magnitude, due in part to their common solar cycle dependence \cite[e.g.][]{ZhaoEA18,BurgerEA22}. Nevertheless, the present approach is taken as the desired result is simply a measure of all potential variations of the pMFPs given the observationally-constrained ensemble of input parameters.

\begin{figure*}
\centering
    \includegraphics[width=0.95\linewidth]{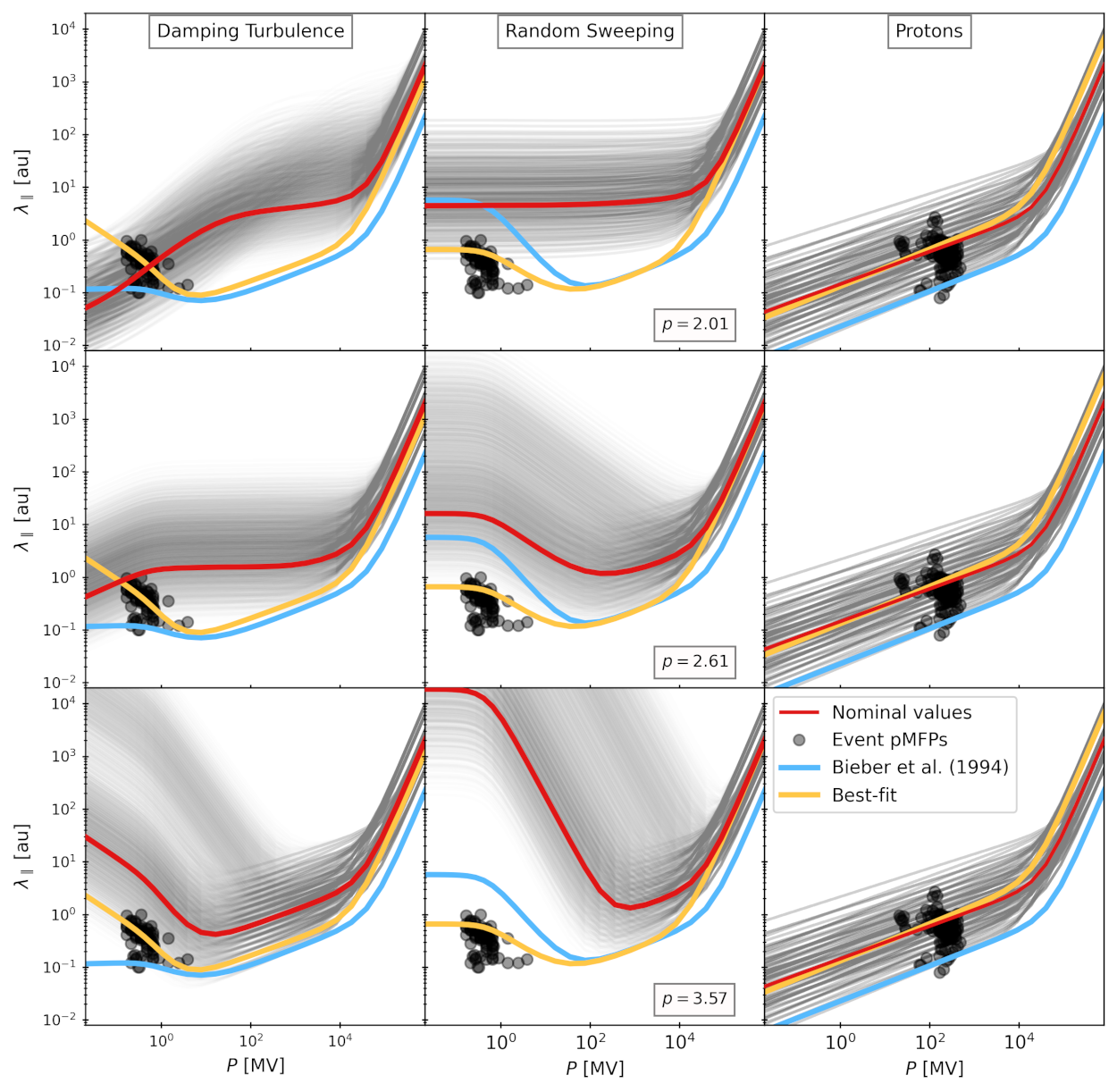}
    \caption{Illustration of every possible iteration of the eight variables $\pm$ the ranges considered, which are used to calculate the pMFP from the dynamical turbulence models, and listed in Table~\ref{tab:1au_params}. The electron damping turbulence pMFP is shown in the left column, the electron random sweeping pMFP is in the center column and the proton pMFP is in the right column, all as a function of rigidity. Each row displays the outcome with a set value of $p$ due to its strong influence on the rigidity dependence of the electron pMFPs. The circles are the results of fitting the events in Table~\ref{tab:events_fitted}. The red line indicates pMFPs calculated using the nominal values and the blue line indicates those calculated using the values originally presented in \citet{BieberEA94}. The pMFPs calculated using a set of 1~au parameters that best fit the observed electron {and proton} data (listed in Table~\ref{tab:1au_params_best}) are also plotted as yellow lines for the DT and RS models, {as well as for protons, where $\mathcal{K}=0$ in Equations~\ref{eqn:turb_mfp_rs} and~\ref{eqn:turb_mfp_dt}}.}
    \label{fig:EldritchHorrors}
\end{figure*} 

The results of these calculations are shown in Figure~\ref{fig:EldritchHorrors}, which displays pMFPs as a function of rigidity for electrons calculated assuming the DT and RS models of dynamical turbulence (left and middle panels, respectively) as well as for protons (right panel), alongside the results from fitting the events in Table~\ref{tab:events_fitted}.
Red lines indicate pMFPs calculated using only the nominal reported values of the various parameters employed here (found in Column 3 of Table~\ref{tab:1au_params}), and blue lines are pMFPs calculated using values employed by \citet{BieberEA94}. 
{The best-fitting lines computed for all the electron and proton event pMFPs are denoted with the yellow lines.} 
Note that Figure~\ref{fig:EldritchHorrors} is ordered by the input value employed for the dissipation range spectral index $p$. 
This is motivated by the fact that this parameter plays a central role in the low energy electron pMFP rigidity dependence \cite[see, e.g.,][]{EngelbrechtBurger13}. 
For instance, the electron RS pMFP at low rigidities displays a $\sim P^{2-p}$ dependence. 
However, this dependence is not so easily expressed for the DT pMFP, due to the Hypergeometric function found in Equation~\ref{eqn:turb_mfp_dt}, and can be seen in Fig.~\ref{fig:EldritchHorrors} to differ considerably between the dynamical turbulence models for a given value of $p$.
It is immediately apparent that all pMFPs shown vary significantly, particularly below $\sim 10^4$~MV, over several orders of magnitude. 
This is particularly true for the electron pMFPs, which can vary up to $5$ orders of magnitude at the lowest rigidities shown. 
At higher rigidities this variation is less significant, spanning only about one order of magnitude, due to the reduced dependence of Term~1 in Equations~\ref{eqn:turb_mfp}~-~\ref{eqn:turb_mfp_dt} on multiple turbulence-related parameters. 
The observational points ({black} circles) shown in Figure~\ref{fig:EldritchHorrors}, which themselves vary over an order of magnitude, fall neatly within the range of pMFPs computed for electrons using the DT model when $p=2.01$, and at the bottom of the range for larger values of $p$. Intriguingly, the lowest reported electron pMFP observations fall below the lowest computed RS pMFPs, more so when the dissipation range is assumed to be steeper, implying that electron pMFPs computed using this dynamical turbulence model can potentially overestimate this quantity. 
This behavior can also be seen in the pMFPs calculated using nominal parameter values ({red} lines), with the RS pMFP again remaining above all the observational points shown, similar to what was reported by \citet{Engelbrecht_2022}. It should be noted, however, that best-fit lines in reasonable to good agreement with observations can be found for both the RS and DT model MFPs ({yellow lines}), albeit for different input parameters (particularly those associated with the dissipation range), listed in Table~\ref{tab:1au_params_best}. Computed proton pMFPs fall well within the range of the observational points. {For protons, the observationally-derived pMFP span almost two orders of magnitude, and do not display a clear rigidity dependence. Interestingly, the best-fit and nominal lines (yellow and red, respectively) are very similar. It should be noted, however, that the theoretical trends shown in Fig.~\ref{fig:EldritchHorrors} cannot necessarily explain the trends implied by some of the observations shown in Fig~\ref{fig:mfpVrig}: a higher trend is required to explain point "G" for pick-up protons, while a lower trend is required to explain points "D" and "+" for SEP protons and GLE events, respectively.}


\begin{deluxetable}{CCCCc}
\tabletypesize{\scriptsize}
\tablewidth{\columnwidth} 
\tablecaption{The {best-fitting turbulence parameter} values (for both the DT and RS models for electrons, and for protons) to the inverted event pMFPs, seen in the yellow lines in Figure~\ref{fig:EldritchHorrors}.  \label{tab:1au_params_best}}
\tablehead{
\colhead{Symbol} & \colhead{DT}  & \colhead{RS} & \colhead{Protons} & \colhead{Units}
} 
\startdata 
B_0 & 3.97 & 3.97 & 3.97 & nT \\
\delta B_{slab}^2  & 5.33 & 5.33 & 1.03 & nT$^2$ \\
s & 1.65 & 1.65 & 1.65 & -- \\
p & 3.57 & 2.61 & 3.57 & -- \\
k_d & 8.44 & 8.44 & 0.12 & $\times10^{-3} \text{~km}^{-1}$ \\
k_{min} & 2.28 & 3.50 & 2.28 & $\times10^{-7} \text{~km}^{-1}$\\ 
V_A & 56.77 & 79.36 & 79.36 & km~s$^{-1}$ \\
\alpha_D & 0.5 & 1.0 & 0.5 & -- \\
\enddata

\end{deluxetable} 

\section{Discussion}

In this paper, we selected a number of magnetically well-connected SEP events measured near $1$~au and compared the intensity-time profiles to results from a 1D SEP propagation model to determine the optimal set of transport parameters. In comparison to previous studies, we incorporated several newly developed tools to increase the accuracy of fitting predictive model outcomes to observations, improving on the `by-eye' methods commonly used in the past, so finding new observational values for electron and proton pMFPs for a broad range of rigidities at $1$~au. The resulting pMFPs are compared to previous estimates from various studies, and to existing theoretical models. The inversion results for the pMFP are consistent with previous work, displaying considerable inter-event variations also seen in the results of prior studies.

We construct an observationally-based ensemble of values of turbulence quantities at 1~au in order to use these as inputs to evaluate theoretical expressions for the pMFP employed in previous electron and proton modulation studies derived assuming either the random sweeping or damping models of dynamical turbulence. This allows us to quantify the range within which $\lambda_{\parallel}$ is expected to vary based on theoretical considerations. The significant pMFP variations in the inversion result and in the theoretical expressions, as seen in Figure~\ref{fig:EldritchHorrors}, demonstrate the sensitivity of electron and proton transport coefficients to differing heliospheric plasma conditions, implying that the change in transport conditions naturally occurring in the inner heliosphere may be the underlying reason behind the fairly large variations in pMFPs reported in studies such as this one. While these variations are mostly consistent with the theoretical expectation, we find, in agreement with \citet{Engelbrecht_2022}, that the electron MFPs derived assuming the damping model to agree best with observations, and that the electron pMFP is extremely sensitive to the onset and shape of the dissipation range of the turbulence spectrum.

{The considerable inter-event variation of the pMFPs reported on in this study, coupled with the large variations in theoretical pMFPs calculated assuming an observationally-motivated ensemble of turbulence input parameters, for both protons and electrons, has significant consequences for particle transport modelling in the broader heliophysical context. These results point directly to the need for a more nuanced modelling of the transport coefficients assumed in such models than is often currently done, regardless of whether solar energetic particle, Jovian electron, or galactic cosmic ray transport is considered. In order to model particular particle observations, such transport modelling would need to carefully take into account observed turbulence conditions spanning the time intervals relevant to the transport of the particles in question \cite[for example, over a span of approximately a year for galactic cosmic rays, see, e.g.,][]{Florinski09,Straussetal2011,wang,kol}, which in turn also implies taking into account the broader spatio-temporal variations of these parameters. Although preliminary studies of this type have been made \cite[e.g.][for galactic cosmic ray transport]{EngelbrechtMoloto21}, there is still considerable scope for refinement, not only in terms of the need for more comprehensive modelling of turbulence and larger-scale plasma parameters, but also of the diffusion coefficients of these particles \citep{EngelbrechtEA22b}. The study of the transport of solar energetic particles \cite[e.g.][]{Straussetal2017a,vandenBergEA20,jabus21}, as well as of Jovian electrons \cite[see][]{StraussEA24}, would also benefit from such an approach. This is, for example, particularly true where the transport of low-energy electrons is considered, given the dependence of these particles' pMFPs on dissipation range turbulence parameters that are observed to vary considerably. It should also be noted that it can reasonably be expected that the amount of variation reported on here for the parallel MFP could also be seen in the perpendicular MFPs of the particles considered here, given that theoretical expressions derived from different scattering theories \cite[for a review of these, see][and references therein]{EngelbrechtEA22b} also depend on significantly-varying turbulence input parameters. This latter point highlights the need for higher-dimensional modelling approaches such as that presented here, taking into account at the least 2D particle transport effects, in order to acquire observational estimates for the perpendicular MFP as well as the pMFP. To date, relatively few studies have done so, and have often been limited to the study of Jovian electrons \cite[e.g.][]{ZhangEA07,Vogtetal2020,Engelbrecht_2022}, with relatively few studies of solar energetic particle transport \cite[e.g.][]{drogeEA14, Straussetal2017a}, motivated by the additional complexity of such a modelling approach. This will be the subject of future work.
}

{Lastly, it should be noted that} the unavailability of observed anisotropy data from several instruments restricts the current study to focusing only on the late isotropic phase of the SEP events and therefore only on the derived pMFP values. If anisotropy measurements are available, it would be possible to also derive the associated acceleration and/or injection time profiles ($\tau_a$, $\tau_e$) of SEPs close to the Sun. This underscores the need for accurate SEP anisotropy measurements for both electron and proton populations. Fortunately, such anisotropy measurements are currently being made by the Solar Orbiter and Parker Solar Probe spacecraft, both exploring the very inner heliosphere, and will be utilized in future versions of this work.
\acknowledgments

This work is based on the research supported in part by the National Research Foundation of South Africa (NRF grant numbers RA170929263913, SRUG220322419, 142149, and 137793). Opinions expressed and conclusions arrived at are those of the authors and are not necessarily to be attributed to the NRF. The responsibility of the contents of this work is with the authors. JTL acknowledges partial financial support from the South African National Astrophysics and Space Science Program (NASSP) and the North-West University (NWU). Figures prepared with Matplotlib \citep{Hunter:2007} and certain calculations done with NumPy \citep{Harrisetal2020}. ND acknowledges the support of the Academy of Finland (SHOCKSEE, grant number 346902). DR acknowledges financial support from the National Science and Technology Development Agency (NSTDA) and National Research Council of Thailand (NRCT): High-Potential Research Team Grant Program (N42A650868) and from the NSRF via the Program
Management Unit for Human Resources \& Institutional Development, Research and Innovation
(B39G670013). {The SERPENTINE project was funded by the European Union’s Horizon 2020 framework program.}

\bibliography{ref}

\end{document}